\documentclass[10pt,aps,pra,notitlepage,twocolumn]{revtex4-1}
\usepackage{xcolor}
\usepackage[bookmarks=false,colorlinks=true,allcolors = blue, citecolor = red]{hyperref}
\usepackage{amsmath,amssymb}
\usepackage{bm}
\usepackage{physics}
\usepackage{graphicx}
\usepackage{ytableau}

\usepackage{natbib}
\newcommand{\sfrac}[2]{{\textstyle\frac{#1}{#2}}}
\newcommand{\yn}{\mathbb{Y}}

\newcommand{\kett}[1]{\ket{#1}\!}
\newcommand{\up}{\uparrow}
\newcommand{\dw}{\downarrow}

\newcommand{\aref}[2]{\hypersetup{urlcolor=violet}\href{#1}{#2}\hypersetup{urlcolor=blue}}

\begin{document}

\title{Lowest energy states of  an $O(N)$ fermionic chain}
\author{Tigran Hakobyan}
\email{tigran.hakobyan@ysu.am}
\email{tigran.s.hakobyan@gmail.com}
\affiliation{Yerevan State University, 1 Alex Manoogian Street, Yerevan, 0025, Armenia}

\begin{abstract}
A quite general finite-size chain of fermions  with $N$ internal degrees of freedom (flavors)  and
$O(N)$ symmetry  is considered.
In the case of the free boundary condition, we prove that the ground state in the invariant sector having exactly $m$ flavors
with an odd particle number is represented by a single rank-$m$ antisymmetric multiplet. For the even-length chains, its particle-hole
quantum number (if it's a good one) is given by the parity of the $m$.
For the odd-length chains, the particle-hole symmetry leads
to the twofold degeneracy among the conjugate multiplets. Similar statements are proven
for the $O(N)$ mixed-spin  chains in antisymmetric representations. The results are extended to the
long-range interacting fermions and (partially) to the translation invariant chains.
\end{abstract}

\maketitle

\section{Introduction}

The degeneracy and quantum numbers of the ground state
have an important bearing on the  low-temperature behavior of quantum
systems. Apart from various numerical and approximate approaches,
there are certain explicit methods to reveal them
for the spin and fermion lattice systems. 
One such approach is based on the existence and properties of the
 basis where all off-diagonal elements of the Hamiltonian's matrix take
 nonpositive values (nonpositive basis).
As a consequence,  the ground state of the spin-$\frac12$
translation invariant antiferromagnetic Heisenberg  model
with an even number of sites is a unique spin singlet \cite{M55,LSM61}.
A simple structure  of the classical ground state case (Neel state),
is not retained in the quantum case.
 However, the quantum ground state inherits certain properties
 from its classical counterpart like  the degeneracy degree and spin value.
Using the structure of the $SU(2)$ spin multiplets, this property
was extended to  antiferromagnetic systems with arbitrary spins on bipartite lattices  \cite{LM62},
repulsive Hubbard \cite{Lieb89} and  periodic Anderson \cite{Ueda92} models at half filling.
 Similar features  was established for  a more common
 class of the $SU(2)$ invariant fermionic chains \cite{Amb92}.

The  extension to the $SU(N)$ symmetric spin  and fermionic
 chains was formulated and proven too \cite{AL86,sorella96,Li01,H04,Li04,H10}.
  In some cases, the uniqueness of the lowest level multiplets in the sectors with
 fixed total spin values and the antiferromagnetic ordering of related energies
 was  established  \cite{LM62,Amb92,H04,H10}.
Higher symmetries may emerge
at special values of the parameters, in the case of   orbital degeneracy \cite{Li98},
as well as at the quantum critical point in  the low-energy limit.
One can mention in this respect the $SO(5)$ symmetry
unifying the antiferromagnetism and high-temperature  superconductivity
\cite{Zhang} (for a review, see Ref.~\cite{so5}).
Moreover, experimental capacities now
enable  to fabricate and  control the artificial quantum systems based on
 ultracold atoms trapped in optical lattices. In particular,
the fermionic alkaline earth atoms realize quantum models
possessing the
unitary   symmetry~\cite{Honercamp04}
(see Ref.~\cite{rey14} for a review).

In this paper, we study a finite-size  chain of interacting fermions
endowed with $N$ internal degrees of freedom (spins or flavors).
The model is defined in terms of the usual (complex)  and
Majorana (real) fermions. We take advantage of both formulations.
Note also that the second representation is actual
due to the recent interest in interacting Majorana fermions \cite{inter-maj}.
The Hamiltonian
remains invariant with respect to the $O(N)$ rotations in the flavor space
(including improper ones).
It has  quite general  multi-fermion interactions.
In contrast to its $U(N)$ invariant counterpart, the system does not
preserve the total number of particles with a given flavor. Instead,
it keeps the related parities.

The parity operators constitute a discrete subgroup
 of reflections with respect to the flavor directions, $Z_2^{\times N}$.
Their eigenvalues $\sigma=\pm 1$ (even/odd) define the  invariant subspaces
of the Hamiltonian.  Such subspaces have equal dimensions and can be mapped to each other by
the Majorana fermion operators. Moreover, the $\sigma$ subspaces with the same
number of odd flavors, $m$, are degenerate and combined into a single
invariant sector.


For a wide range  of  coupling constants, we prove that the lowest
energy $O(N)$ multiplet  in any such $m$ sector is unique and
represented by an $m$th-order antisymmetric tensor.
The components form the nondegenerate lowest energy states
(the relative ground states) in the corresponding $\sigma$ subspaces.
Thus, the ground state in  the $m=0$ sector (where all parities are
even)  is a unique $O(N)$ singlet. At the same time,   in  the $m=N$ sector
(where the parities are odd), it is a unique  pseudoscalar
(i.e. behaves as a singlet under proper rotations while
changing signs under improper ones).
An additional degeneracy is not banned and may happen
 for special values of couplings with accidental symmetries.
 In particular, in the limit of the $N$ decoupled Kitaev chains \cite{kitaev01},
 the total ground state completely breaks the $Z_2^{\times N}$
 symmetry.

 We also consider Hamitonians with  particle-hole symmetry.
 The related $Z_2$ group commutes  with the $SO(N)$ symmetry.
 The impact on the spectrum depends on the parity of the chain's size.
  For the even-length
 chains, it is consistent with the whole $O(N)$ symmetry, including
 improper rotations. The lowest energy states acquire a
 particle-hole quantum number given by the parity of $m$.
 For the odd chains, this map alters all parities, $\sigma\to -\sigma$, which
 leads to an additional twofold degeneracy.

We also examine  in the same context the $O(N)$ mixed-spin chains in the
antisymmetric representations.
They emerge at the particular limiting values
of the parameters when the on-site particle numbers become persistent.
The total parity turns into a constant, dependent on the chain's size,
so the independent reflection generators form a $Z_2^{\times (N-1)}$ group.
It is argued, however, that the aforementioned  results on the uniqueness
and $O(N)$ structure of the relative ground states
 remain valid for the spin chains too.
The results extend our previous studies of the bilinear-biquadratic Heisenberg model
with spins in the vector representation   \cite{H15}.

The long-range interactions may also be involved into the fermionic
 Hamiltonian in a way to maintain the above properties of the
 nearest-neighboring chain. The distant interaction contains
 an additional sign-valued tail depending on the intermediate
 fermions.
Finally, we show that for the translation-invariant chain,
 the lowest energy state in the odd-parity sector
has zero momentum.

The article is organized as follows. In Sec.~\ref{sec:def},
we describe in detail the model and its symmetries in terms of the standard
and Majorana fermions.
In Sec.~\ref{sec:lowest},  the properties of the invariant
 subspaces and sectors are described. Then the basis, where all off-diagonal elements
of the Hamiltonian  are nonpositive, is presented using  the standard
and Majorana fermions, as well as hard-core bosons. Finally,
 the aforementioned result about
the $O(N)$ structure of the lowest energy states is proven.
In Sec.~\ref{sec:ph},  this result is extended
to fermionic chains  with  particle-hole symmetry.
Section~\ref{sec:further} is devoted  to the fermionic chains with
 long-range interactions, translation invariance, and
mixed-spin chains in the antisymmetric representation.
Finally, in the Appendixes, we derive the complete spectrum and multiplet
structure of the two-site system with two and three flavors.

\section{$\boldsymbol{O(N)}$ symmetric fermionic chain}
\label{sec:def}
\subsection{Standard fermions}
Consider the  extended Hubbard chain of length $L$ described by the Hamiltonian
\begin{widetext}
\begin{equation}
\label{H}
\begin{aligned}
\hat H=&-\sum_{x,a}\big( t_xc^+_{x+1,a}c_{x,a}  + r_x c_{x+1,a}c_{x,a}\,+\text{H.c.}\big)
-\sum_{x,a, b} \Big( f_x c^+_{x+1, b} c^+_{x,a} c_{x+1,a} c_{x, b}
- g_x c^+_{x+1,a} c^+_{x,a} c_{x+1, b} c_{x, b}
\\
&+ h_x c^+_{x+1, a} c^+_{x+1,b} c_{x,b} c_{x, a}
+e_x c_{x+1, a} c_{x+1,b} c_{x,b} c_{x, a} +\text{H.c.}\Big)+  V(n_1,\dots,n_L).
\end{aligned}
\end{equation}
\end{widetext}
The open boundary conditions
are supposed so  the position index in the sums, $x$,
varies from $1$ to $L-1$.
There are $N$ different species (flavors) of fermions, which are labeled by $a,b$.
The creation-annihilation    operators $c^\pm_{x,a}$
obey the standard  anticommutation relations.

The potential $V$ depends on the local fermion occupation numbers:
$$
n_x=\sum_a c_{x,a}^+c_{x,a}.
$$
Its explicit form does not matter
here.
The Hubbard potential, $V=\sum_x n_x^2$, is a particular case.

The coupling coefficients in the Hamiltonian depend on the fermion position.
In this paper, we will set them positive. More explicitly, we impose
\begin{equation}
\label{pos}
t_x,r_x>0,
\qquad
f_x,g_x,h_x,e_x\ge 0.
\end{equation}
These conditions may be even weakened, see Eq.~\eqref{rt} below.

The $t$ term in the Hamiltonian describes the single fermion hopping between
neighboring sites.  The $r$ term is responsible for the creation-annihilation of
the superconducting fermion pairs of same flavor. The remaining part of the Hamiltonian
is responsible for the four-fermion interaction.
The $f$ term  swaps the fermions with different flavors on adjacent sites,
$| a b\rangle\to|b a\rangle$.
 The $g$ term
replaces a pair of adjacent fermions of a same type  with an other-type pair,
$| a a\rangle\to|b b\rangle$. The $h$ term moves a fermion pair providing the system
with pair-hopping opportunity. Finally, the $e$ term
creates and annihilates four neighboring particles, two per node.

For $N=1$ with a single fermion per site, the pair-hopping term disappears. The remaining two four-fermion
interactions are reduced to the density-density interaction between adjacent sites, which may be
included in the potential:
\begin{equation}
\label{Hone}
\begin{aligned}
\hat H=&-\sum_x\left( t_xc^+_{x+1}c_{x}  + r_x c_{x+1}c_{x}+\text{H.c.}\right) +  V+\delta V,
\\
&\delta V=2\sum_x (f_x-g_x) n_{x+1}n_x.
\end{aligned}
\end{equation}
For the special case when $V+\delta V$ is set to the chemical potential, the system can be
considered as a local analog of the Kitaev chain \cite{kitaev01}.

For $N>1$, the Hamiltonian \eqref{H} is invariant under the global $SO(N)$ rotations,
\begin{align}
\label{Lab}
[\hat H,\hat{L}_{ a b}]=0,
\qquad
\hat{L}_{ a b} =\sum_x L^{ a b}_x,
\end{align}
where the local rotations are provided by the generators:
\begin{equation}
\label{Lab'}
L_x^{ a b}=\imath(c^+_{x,a}c_{x,b}-c^+_{x,b}c_{x,a}).
\end{equation}

Neither the number of particles with a given species,
\begin{equation}
\label{na}
\hat n_a=\sum_x c_{x,a}^+c_{x,a},
\end{equation}
 nor the total number of particles
is conserved in the system.
Instead, as is easy to see, the related  parities are good quantum numbers.
Denote by $\hat \sigma_a$ an  operator describing the
parity of the number of fermion with  flavor  $a$:
\begin{equation}
\label{parity}
[\hat H, \hat\sigma_ a ]=0,
\qquad
\hat\sigma_ a = (-1)^{ \hat n_ a}.
\end{equation}

Such reflections generate  the $Z_2^{\times N}$ group. Together
with the continuous rotations, they make up the orthogonal group $O(N)$ composing
the internal symmetry of the fermionic chain \eqref{H}.

Note that the product of two distinct reflections
define a $\pi$ rotation in a plane inside the flavor space:
$$
\hat\sigma_ a\hat\sigma_{ b}=e^{i\pi \hat{L}^{ a b}}.
$$
Such rotations  form together a subgroup  $Z_2^{\times (N-1)}$.

According to the Pauli exclusion principle,  every site can be occupied by at most $N$ fermions.
There are $2^N$ such states.
The one-particle states, $c^+_{a}|0\rangle$, form the defining
representation of $O(N)$.
Due to the Fermi-Dirac statistics,
the multi-particle states,
\begin{equation}
\label{Yk}
c^+_{a_1}\cdots c^+_{a_m}|0\rangle,
\end{equation}
comprise the $\binom{N}{m}$-dimensional antisymmetric multiplet.
The empty and completely filled nodes are, respectively,  a scalar
and pseudoscalar.

The $SO(N)$ structure of the single-node states is trickier \cite{Hamermesh}.
The  conjugate multiplets with $m$ and $N-m$ fermions become equivalent.
Moreover, in the case of two flavors, $N=2$, the group $SO(2)$ is Abelian. Then
the single-particle representation is reducible and splits into
the symmetric and antisymmetric combinations.

At  the limiting point where three couplings  vanish, $r_x=g_x=e_x=0$, the symmetry
is expanded to the unitary group $U(N)$. The additional generators are provided by
the symmetrized bilinear components:
\begin{equation}
\label{Tab}
\hat{T}_{ a b} =\sum_x T^{ a b}_x,
\qquad
T_x^{ a b}=c^+_{x,a}c_{x,b}+c^+_{x,b}c_{x,a}.
\end{equation}
The diagonal part consists of the fermion number operators, $\hat n_a=\frac12\hat T_{aa}$, which
are preserved in this case.

To reveal the structure of the Hamiltonian \eqref{H}, let us
present it in the form
\begin{equation}
\label{Hoff}
\hat H=\sum_{x=1}^{L-1} \left(H_{x+1\,x} + H_{x+1\,x}^+ \right)+ V'(n_1,\dots,n_L)
\end{equation}
with the slightly modified potential:
\begin{equation}
\label{V'}
 V'=V+2\sum_{x=1}^{L-1} f_x n_{x+1}.
\end{equation}
Next, present  the local Hamiltonian as follows (we set  $y=x+1$
to shorten the formula):
\begin{equation}
\label{Hyx}
\begin{aligned}
H_{yx}  =&-t_x  K_{yx} - r_x P_{yx}
        - f_x K_{yx}K_{xy} - g_x  P_{yx}^+P_{yx}
        \\
        &-h_x K_{yx}^2-e_x P_{yx}^2.
\end{aligned}
\end{equation}
Here we have introduced the two  $O(N)$ invariant bilinear combinations
 of fermionic operators:
 \begin{equation}
 \label{KP}
 \begin{gathered}
  K_{yx}=\sum_a c^+_{y,a}c_{x,a}=\bm{c}_y^+\cdot \bm{c}_x,
\\
P_{yx}=\sum_a c_{y,a}c_{x,a}=\bm{c}_y\cdot \bm{c}_x.
\end{gathered}
\end{equation}
Notice that the $K_{yx}$ possesses   a larger,  $U(N)$, symmetry  while $P_{yx}$ does not.

  The equivalence of the representations \eqref{H} and
 \eqref{Hoff} is easy to establish using the canonical anticommutation
 relations.
Note that the operators \eqref{KP} obey the conditions $K_{yx}^+=K_{xy}$ and $P_{xy}=-P_{yx}$, as well as
the following commutation rules:
\begin{equation}
\label{comKK}
\begin{gathered}
{[}K_{xy},K_{yx}]=n_x-n_y,
\quad
[P_{yx}^+,P_{yx}]=n_x+n_y,
\\
[K_{xy},P_{xy}]=0.
\end{gathered}
\end{equation}

The usual Heisenberg interaction between neighboring $SO(N)$ spins is expressed via them as
\begin{equation}
\label{spin-spin}
\bm{L}_y\cdot \bm{L}_x=\sum_{a<b}L_y^{ab}L_x^{ab}=-K_{yx}^+K_{yx}-P_{yx}^+P_{yx}+n_x.
\\
\end{equation}
The local fermion number, $n_x$,  may be added to the potential and will not be essential
 in the current context.
Notice that the invariance of the right side under the coordinate replacement $x\leftrightarrow y$
follows from the commutation relation \eqref{comKK}.
The spin-exchange term appears in the Hamiltonian, for instance, when the parameters obey
the condition $f_x=g_x$.

\subsection{Majorana fermions}

In this section, the initial Hamiltonian \eqref{H} is represented as an $O(N)$ chain
 of interacting Majorana fermions.

It is well known that a single complex fermion is  equivalent to a pair of
real, or Majorana fermions.
The relation among both  representations is provided by the
 map \cite{kitaev01},
\begin{equation}
\label{gamma}
c_{x,a}^\pm=\frac{\gamma_{x,a}^{(1)}\mp\imath \gamma_{x,a}^{(2)}}{2},
\end{equation}
and its inverse:
\begin{equation}
\label{inv}
\gamma_{x,a}^{(1)}=c^+_{x,a}+c_{x,a},
\qquad
\gamma_{x,a}^{(2)}=\imath(c^+_{x,a}-c_{x,a}).
\end{equation}
The Majorana fermions are identical to
their own antiparticles and described by the Hermitian unitary operators  $\gamma_{x,a}^{(\lambda)}$,
with the upper index $\lambda=1,2$ separating individual particles in the pair. These particles
have become quite popular recently, see  Ref.~\cite{wil09} for a short review on the subject.

The Majorana operators generate the $2NL$-dimensional Clifford algebra:
$$
\big\{\gamma_{x,a}^{(\lambda)},\gamma_{y,b}^{(\lambda')}\big\}=2\delta_{ab}\delta_{xy}\delta_{\lambda\lambda'}.
$$
The  number of $a$-type on-site fermions and its parity  can be
expressed via them:
\begin{equation}
\label{na-major}
2n_{x,a}-1=\imath\gamma_{x,a}^{(1)}\gamma_{x,a}^{(2)},
\qquad
\sigma_{x,a}=-\imath\gamma_{x,a}^{(1)}\gamma_{x,a}^{(2)}.
\end{equation}
As a result, the overall parity \eqref{parity} is just a product of all Majorana operators with a certain
phase factor ensuring the involutivity  \cite{kitaev11}:
\begin{equation}
\label{chiral}
\hat\sigma_a=(-\imath)^L\prod_{x=1}^L\gamma_{x,a}^{(1)}\gamma_{x,a}^{(2)}.
\end{equation}
It is worth mentioning that in the Dirac matrix context, it corresponds to the \emph{chiral}
gamma-matrix which anticommutes with all $\gamma$-s of the same type:
\begin{equation}
\label{gamma5}
\hat\sigma_a=\gamma_{2L+1,a}.
\end{equation}
The right/left chirality sectors then correspond to the states with even/odd parities
respectively.

The local symmetries \eqref{Lab} and \eqref{Tab} can be also
expressed in terms of Majorana fermions. The rotation
generators in this form are known from the spinor representation of the orthogonal group:
\begin{equation}
\label{LabM}
L_x^{ab}=\frac\imath4\sum_{\lambda=1,2}\big[\gamma_{x,a}^{(\lambda)},\gamma_{x,b}^{(\lambda)}\big].
\end{equation}

The local building blocks of the Hamiltonian are expressed through the Majorana operators in the following
way:
\begin{multline}
\label{Kmaj}
K_{yx}=
\frac14\sum_{a}\left(
\gamma^{(1)}_{y,a}\gamma^{(1)}_{x,a}+\gamma^{(2)}_{y,a}\gamma^{(2)}_{x,a}
\right)
\\
+\frac\imath4\sum_{a}\left(
\gamma^{(1)}_{y,a}\gamma^{(2)}_{x,a}-\gamma^{(2)}_{y,a}\gamma^{(1)}_{x,a}
\right),
\end{multline}
\vspace{-5mm}
\begin{multline}
\label{Pmaj}
P_{yx}=
\frac14\sum_{a}\left(
\gamma^{(1)}_{y,a}\gamma^{(1)}_{x,a}-\gamma^{(2)}_{y,a}\gamma^{(2)}_{x,a}
\right)
\\
+\frac\imath4\sum_{a}\left(
\gamma^{(1)}_{y,a}\gamma^{(2)}_{x,a}+\gamma^{(2)}_{y,a}\gamma^{(1)}_{x,a}
\right).
\end{multline}

%

In both expressions, the first sum is antisymmetric under the exchange of the coordinates $x$ and $y$. Hence,
it disappears in the double-fermion part of the Hamiltonian.
In contrast, the second sum is symmetric and participates there.

 In Majorana representation the fermionic Hamiltonian \eqref{Hoff} acquires the following explicit
 form:
\begin{widetext}
\begin{equation}
\label{HMaj}
\begin{aligned}
\hat H=&-\frac{\imath}{2}
\sum_{x,a}\left[ (t_x+r_x) \gamma^{(1)}_{x+1,a}\gamma^{(2)}_{x,a}  - (t_x-r_x) \gamma^{(2)}_{x+1,a}\gamma^{(1)}_{x,a}\right]+V+\delta V
 \\
&+\frac18\sum_{x,a, b}\sum_{\lambda=1,2}\left[ (f_x+g_x-h_x-e_x)
\gamma^{(\lambda)}_{x+1,a}\gamma^{(\lambda)}_{x,a} \gamma^{(\lambda)}_{x+1,b}\gamma^{(\lambda)}_{x,b}
+(f_x+g_x+h_x+e_x) \gamma^{(\lambda)}_{x+1,a}\gamma^{(\bar\lambda)}_{x,a} \gamma^{(\lambda)}_{x+1,b}\gamma^{(\bar\lambda)}_{x,b}
\right.
\\
&\qquad\qquad\;\; \left.
+( -f_x+g_x-h_x+e_x) \gamma^{(\lambda)}_{x+1,a}\gamma^{(\bar\lambda)}_{x,a} \gamma^{(\bar\lambda)}_{x+1,b}\gamma^{(\lambda)}_{x,b}
+ ( f_x-g_x-h_x+e_x)\gamma^{(\lambda)}_{x+1,a}\gamma^{(\lambda)}_{x,a} \gamma^{(\bar\lambda)}_{x+1,b}\gamma^{(\bar\lambda)}_{x,b}
 \right],
\end{aligned}
\end{equation}
\end{widetext}
where the bar over $\lambda=1,2$ inverts the order of two particles  in the Majorana pair, $\bar\lambda=2,1$.
The potential $V$  depends on the local fermion numbers \eqref{na-major}.

The above Hamiltonian describes $N$ interacting Majorana chains \cite{inter-maj}.
In the absence of the four-fermion  interactions, it describes the $N$ decoupled  chains,
any of which extends the well-known Kitaev model,
describing   tight-binding chains with $p$-wave superconducting pairing \cite{kitaev01},
out of the homogenous point.

Recently, the Majorana representations of conventional lattice fermions has been successfully
applied for elaboration of sign-free  Monte Carlo algorithms
\cite{MC} for studying the ground-state degeneracy of interacting spinless fermions
\cite{Wei15} using the reflection positivity \cite{Lieb89}.
We apply it throughout the current paper, in particular,  to uncover the structure of the invariant subspaces
and the particle-hole map.

\section{Lowest energy multiplets}
\label{sec:lowest}
\subsection{Invariant subspaces}
\label{sub:inv}
In this section, we will describe the subspaces,
which remain invariant under the Hamiltonian's action.

There are $2^{NL}$ different states in the entire  space $\mathcal{V}^L$.
We partition the $\mathcal{V}^L$ into the $2^{N}$ subspaces, each
 characterized by  its own set of
reflection quantum numbers \eqref{parity}:
\begin{equation}
\label{V}
V^L_{\sigma_1\dots\sigma_{N}}=\{ \psi\,|\,\hat\sigma_a\psi=\sigma_a\psi\}.
\end{equation}
We call them $\sigma$ subspaces following an analogy with the spin system \cite{LM62}.
Since the parities are good quantum numbers \eqref{parity}, the   Hamiltonian \eqref{H}
remains invariant in any $\sigma$ subspace.

All such subspaces are mapped to each other by a single Majorata operator
$\gamma_a=\gamma^{(\lambda)}_{x,a}$,
\begin{equation}
\label{gamma-map}
\gamma_a\, V^L_{\sigma_1\dots\sigma_a,\dots\sigma_{N}}
=V^L_{\sigma_1\dots -\sigma_a\dots\sigma_{N}}.
\end{equation}
In this way, any $\sigma$ subspace is obtained from a single one,
for example,
\begin{equation}
\label{V++}
V^L_{\sigma_1\dots\sigma_{N}}=
\prod_{a=1}^N\gamma_a^{\frac12(1-\sigma_a)}\, V^L_{+\dots +}.
\end{equation}
So, they have the same dimension:
\begin{equation}
\label{dim}
\text{dim}\,V^L_{\sigma_1\dots\sigma_{N}}=2^{N(L-1)}.
\end{equation}

Sometimes is more convenient to label the invariant subspaces by the
values of odd flavors,
\begin{equation}
\label{Va}
 V^L_{a_1\dots a_m}
  :=V^L_{\sigma_1\dots\sigma_{N}}
  \quad
  \text{with}
  \quad
  \sigma_a=\prod_{i=1}^m(-1)^{\delta_{a_ia}}.
\end{equation}
Which notation of these two is used will be clear from the context.
The new one depends
on an $m$-combination of the $N$ flavor's set but not on their order. Hence,
it is symmetric on the flavor indexes.

In contrast to the Hamiltonian, the orthogonal symmetry mixes different $\sigma$
subspaces.
Consider the symmetric group
of permutations between the flavors, ${S}_N\subset O(N)$.
It permutes the reflection operators  and the indexes,
\begin{equation}
\label{perm}
s\hat\sigma_ a s^{-1}=\hat\sigma_{s( a)},
\qquad
sV^L_{a_1\dots a_m}  = V^L_{s(a_1)\dots s(a_m)},
\end{equation}
where $s\in {S}_N$, or, equivalently,
\begin{equation}
sV^L_{\sigma_1\dots\sigma_{N}}  = V^L_{\sigma_{s(1)}\dots \sigma_{s(N)}}.
\end{equation}

Due to   this symmetry, the Hamiltonian has the same spectrum
on all  invariant subspaces, which have the same
number $m$ of odd parities,
We unify them into the $\binom{N}{m}$-fold \emph{degenerate sector}
of dimension $2^{N(L-1)}\binom{N}{m}$;
\begin{align}
\label{W}
\mathcal{V}^L_{m}=\bigoplus_{ a_1<\dots<a_m}V^L_{a_1\dots a_m}.
\end{align}
Clearly, the total space of states splits into the sum of all possible sectors:
\begin{equation}
\label{V-dec}
\mathcal{V}^L=\bigoplus_{m=0}^N \mathcal{V}^L_m.
\end{equation}

\subsection{Spectrum of two-site chains}
\label{sub:L=2}
Before making general statements, let us consider a toy system
on two nodes with the Hamiltonian,
\begin{flalign}
\label{H2site}
\begin{aligned}
\hat H  =&-t K- r P -hK^2-e P^2\,+\,\text{H.c.}
    \\
    &- f (K^+ K+K K^+)- g (P^+P+ PP^+),
\end{aligned}
\\
\intertext{with}
 \label{o2KP}
  K=\sum_{a} c^+_{2,a}c_{1,a},
\qquad
P=\sum_{a} c_{2,a}c_{1,a}.
\end{flalign}
It respects the particle-hole inversion
(see Sec.~\ref{sec:ph} below). Together
with the lattice reflection symmetry, this
significantly simplifies the solution.
We will stick with the lowest spins:
$N=2,3$.

\medskip

First, we describe briefly the energy spectrum and the multiplet
structure of  the $O(2)$ model. The detailed
calculations  are done in Appendix~\ref{sec:o2}.
We keep the usual notations and associate the two flavors
 with the spin-up and spin-down states so that  $a=\up,\dw$.
There are three sectors \cite{fnote},
\begin{equation}
\label{o2W}
{\cal V}_0=V,\qquad
{\cal V}_1=V_\up\oplus V_\dw,\qquad
 {\cal V}_2=V_{\up\dw},
\end{equation}
consisting of the four-dimensional invariant
$\sigma$ subspaces \eqref{V}, \eqref{Va} and \eqref{W}:
\begin{equation}
\label{o2Va}
 V=V_{++}, \quad
 V_{\up}=V_{-+},
 \quad
  V_{\dw}=V_{+-},
 \quad
V_{\up\dw}=V_{--}.
\end{equation}

The $16$ independent two-site states  are
 partitioned into  three singlets,
 three pseudo-singlets (describing by the Levi-Civita tensor, $\epsilon_{ab}$),
 four vector doublets, and  a single doublet,
behaving  under rotations
as a symmetric traceless tensor:
\begin{equation}
\label{psiab}
\psi_{ab}=\psi_{ba},
 \qquad
\sum_a\psi_{aa}=0.
\end{equation}
\begin{figure*}[t]
\includegraphics[width=\textwidth]{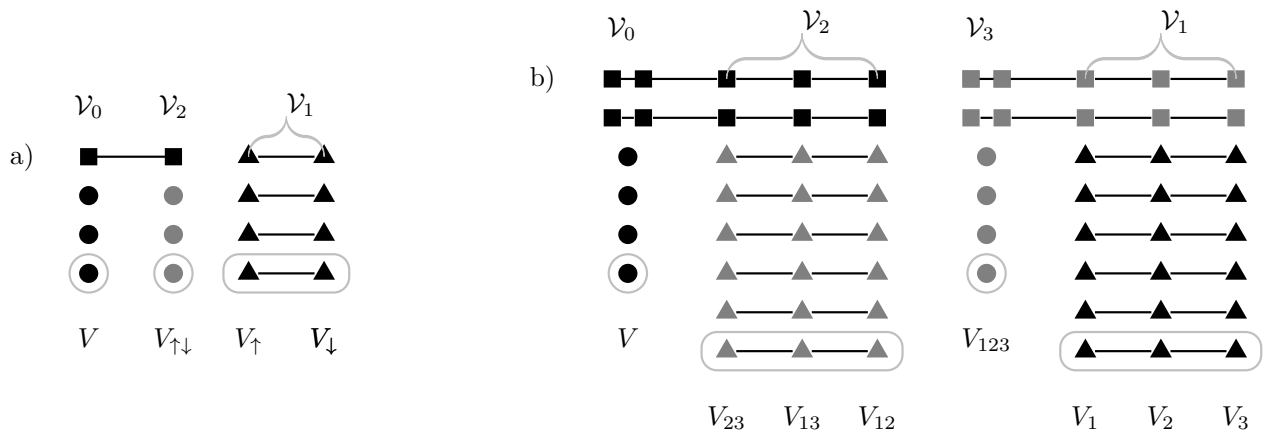}
\caption{\label{fig:o23}The structure of multiplets in the two-site $O(2)$ (a) and $O(3)$  (b)
fermionic chain.
Each column depicts a $\sigma$ subspace $V_{a_1\dots a_m}$ labeled by the odd flavor
values.
The identical columns unite into a single sector, ${\cal V}_m$.
The black circles, triangles, and squares represent
 the singlet, vector, and (traceless) symmetric states, respectively. The gray dots
 in the listed order
denote the pseudo-singlet, pseudo-vector and pseudo-symmetric states. The lines
connect  the elements of  the same multiplet and, hence, are on the same energy level.
Each contour encloses the lowest-energy multiplet
within the corresponding sector.}
\end{figure*}

In Fig.~\ref{fig:o23}(a), which summarizes the results in  Appendix~\ref{sec:o2},
the states of the listed multiplets
are depicted, respectively, by the black and gray circles, the black  triangle, and black square.
A line connects  the elements of the same multiplet,
which, clearly, are on the same energy level.

Each column in Fig.~\ref{fig:o23} represents a certain $\sigma$ subspace.
As we see, the singlets and pseudo-singlets are gathered, respectively,
 in the even, ${\cal V}_0$,
and odd, ${\cal V}_2$, sectors.
At the same time, the vectors compose the remaining sector, ${\cal V}_1$.
 Each vector doublet
spreads along  the two $\sigma$ subspaces with opposite parities,  $V_\up$ and $V_\dw$,
which make up that sector.

Every contour encloses the lowest-energy multiplet
within the corresponding sector. In the allowed parameter's
region, it is unique. Moreover, we see that in the sector
${\cal V}_m$ ($m=0,1,2$)
it is characterized by the $m$th-order antisymmetric
tensor (the scalar, vector, and pseudoscalar, respectively).

Comparing together the minimal energies of individual sectors
[see $E_0$ in the expressions
\eqref{E++}, \eqref{E--}, and \eqref{E+-} in Appendix~\ref{sec:o2}],
we conclude that there is no  lowest  among them
for the common  coupling
parameters \eqref{pos}.
In particular, if the latter  are restricted by the conditions
$g=f$,
$e=h$,
$r=t$,
a degeneracy happens among the sectors ${\cal V}_0$
and  ${\cal V}_2$.
Thus, in general, the total ground state may be a unique scalar, a pseudoscalar,
a vector doublet, or  their superposition.

 \medskip

Go over now to the $O(3)$ model and outline briefly the spectrum properties,
derived in detail in Appendix~\ref{sec:o3}.
The entire space of states, ${\cal V}$,  splits into the four sectors,
composed of the eight-dimensional
$\sigma$ subspaces,
\begin{equation}
\label{o3W}
\begin{aligned}
{\cal V}_0&=V,
&
{\cal V}_1&=V_1\oplus V_2\oplus V_3,
\\
 {\cal V}_3&=V_{123},
 \quad
 &
 {\cal V}_2&=V_{12}\oplus V_{13}\oplus V_{23},
 \end{aligned}
\end{equation}
characterized by the following parities:
\begin{equation}
\label{o3Va}
 \begin{aligned}
 V_1&=V_{-++},
 \quad
&V_2&=V_{+-+},
  \quad
   & V_3&=V_{++-},
\\
V_{12}&=V_{--+},
  &V_{13}&=V_{-+-},
 & V_{23}&=V_{+--},
 \\
  V&=V_{+++},
&V_{123}&=V_{---}.
\end{aligned}
\end{equation}

The $64$-dimensional space ${\cal V}$   is distributed along
the four singlets,   six vectors,
two symmetric traceless tensors \eqref{psiab}, depicted again by the black dots
(cycles, triangles, squares) in Fig.~\ref{fig:o23}(b),
 and  their pseudo-analogs with the same multiplicities, drawn by the gray dots.
 Compared to the $O(2)$ case, the two types of new
 multiplets are listed above: the pseudo-vector, which is equivalent to the antisymmetric  tensor
 $v_{ab}=\sum_c\epsilon_{abc}v_c$, and the pseudo-analog of the symmetric traceless
tensor \eqref{psiab}:
\begin{equation}
\label{psiabc}
\psi_{abc}=\sum_d\epsilon_{abd}\psi_{dc}.
\end{equation}
The latter is more known as a third-order tensor with mixed symmetry \cite{Hamermesh}.
Like the $\psi_{ab}$, it has five independent components.

The  distribution of all
multiplets along the $\sigma$ subspaces and sectors is shown in  Fig.~\ref{fig:o23}(b).
In particular, any quintet described by the tensor $\psi_{ab}$ ($\psi_{abc}$)
has two states from the sector ${\cal V}_0$ (${\cal V}_3$), and three others
from the sector  ${\cal V}_2$ (${\cal V}_1$), one per each $\sigma$ subspace.

The properties of the lowest-level states are similar
to those of the $O(2)$ model.
For a given $m=0,1,2,3$, the lowest-energy multiplet of the sector ${\cal V}_m$
is  a unique $m$th-order antisymmetric
tensor (the scalar, vector, pseudovector, and pseudoscalar, respectively).

Again, the total ground state is not determined for the common values of the couplings.
In particular, in the case of $g=f$, $e=h$ and $r=t$,
a degeneracy emerges  among the multiplets and their pseudo-analogs.

The remaining part of the current section is devoted to the extension and
proof of the above statement to the $O(N)$ fermionic chains of arbitrary length
\eqref{H}.

\subsection{Nonpositive basis}
\label{sec:basis}
Here we pick up  a basis where all nonzero off-diagonal matrix elements of the  Hamiltonian
 become negative. This can be achieved by a specific rearrangement of the fermions
in the standard Fock basis, which results in an additional sign factor \cite{AL86,H10}.
First, let us group together  the fermions with the same flavors and set them
 in ascending order by the coordinate. So, define
\begin{equation}
\label{Ax}
A_{\{x\},a} =c^+_{x_1,a}c^+_{x_2,a}\ldots c^+_{x_{n_a},a},
\qquad
x_1<\dots<x_{n_a},
\end{equation}
where $n_a$ is the total number of particles with  the flavor $a$
in the building state. (In their absence, the above operator is set to unity.)
Then define the basic states in the following way:
\begin{equation}
\label{basis}
A_{\{x^1\},1} A_{\{x^2\},2} \ldots A_{\{x^N\},N} |0\rangle.
\end{equation}

Thus, the particles are
arranged first by the flavor numbers, then
by the coordinates. In other words, they are displaced in the \emph{ascending}
order in their multi-index values
when rewriting  the above state in the
standard way,
\begin{align}
\label{basis-order}
\Psi^{x_1\dots x_n}_{a_1\dots a_n}=c^+_{x_1,a_1}\ldots c^+_{x_n,a_n}|0\rangle
\end{align}
with $n=\sum_a n_a$.
The order is defined as
\begin{equation}
\label{order}
(x,a)<(y,b) \quad
\text{if} \quad
\begin{cases}
a<b,
\\
a=b\quad
\text{and} \quad x<y.
\end{cases}
\end{equation}

In general, the wave function \eqref{basis-order} is not a part of a certain
$O(N)$ multiplet apart from  the case when all fermions are located on a single site \eqref{Yk},
see also Eq.~\eqref{init} below.

Since the potentials $V$, $\delta V$ are diagonal in the constructed basis, the
off-diagonal matrix elements of the chain Hamiltonian are generated exclusively
by algebraic combinations of the local operators $K_{x+1\, x}$, $P_{x+1\, x}$ and their conjugates with
positive coefficients, see \eqref{pos}, \eqref{Hoff},
\eqref{Hyx}, and \eqref{KP}.
Therefore, it is enough to show the positivity
 of the selected basis \eqref{basis} for the $K,P$ operators.

Indeed, due to the Fermi--Dirac statistics, the hopping term  $c^+_{x+1, a}c_{x, a}$ acts
nontrivially solely on the states  with the $a$-type fermion on the $x$th position and without it on the $(x+1)$th one
\cite{H10}.
The resulting action  merely replaces the creation operator $c^+_{x, a}$ by
the annihilation one, $c^+_{x+1, a}$.
Similarly, a pair annihilation term,
$c_{x+1, a}c_{x, a}$, acts
nontrivially on the states where both positions are filled with $ a$-fermions, which are presented in
the basic state \eqref{basis} in reverse order,
$|\dots c^+_{x, a}c^+_{x+1, a}\dots |0\rangle$.
It just eliminates both fermions, producing another basic state without any factor.
Clearly, the Hermitian conjugates of both operators, the backward hopping, $c^+_{x, a}c_{x+1, a}$,
and pair creation, $c^+_{x, a}c^+_{x+1, a}$,
act on the basic states in reverse order. So, all matrix elements of the four considered operators
are either $0$ or $1$. Thus the operators $K_{x+1\,x}$,  $P_{x+1\,x}$ and their
conjugates \eqref{KP} can generate only integral matrix elements from $0$ to $N$.

The specific fermion ordering in the basic wavefunctions \eqref{basis}  has a simple explanation in terms of the
well-known Jordan-Wigner transformation \cite{LSM61,harada14}.
Assigning to each multi-index value the three Pauli matrices,  $\tau^\pm,\tau^3$,
we get the system of hard-core bosons. Such particles behave alike fermions (bosons)
at the same (different) points.
Once we have set the ordering \eqref{order},
the fermions and bosons are related by the Jordan-Wigner transformation:
\begin{equation}
\label{JW}
\tau^\pm_{x,a}=c^\pm_{x,a}\prod_{(y,b)<(x,a)}\sigma_{y,b},
\qquad
\tau^3_{x,a}=-\sigma_{x,a}.
\end{equation}
The building blocks of the Hamiltonian \eqref{H},  expressed  in terms
of the Pauli matrices, retain their structure:
\begin{equation}
\label{block}
c^\pm_{x+1,a}c_{x,a}=-\tau^\pm_{x+1,a}\tau^3_{x,a}\tau^-_{x,a}=\tau^\pm_{x+1,a}\tau^-_{x,a}.
\end{equation}
Using the relations \eqref{JW}, it is easy to see that the multi-fermionic wavefunctions \eqref{basis}
in \emph{ascending} order \eqref{basis-order}
can be expressed as  multi-bosonic states:
\begin{equation}
\label{basis-pa}
\Psi^{x_1\dots x_n}_{a_1\dots a_n}
=\tau^+_{x_1,a_1}\ldots \tau^+_{x_n,a_n}
|0\rangle.
\end{equation}
Evidently, the ordering of Pauli matrices on the right side of this equation
is not essential in contrast to the fermion ordering on the left side.
Then the  relations \eqref{Hoff}, \eqref{Hyx}, \eqref{KP}, and \eqref{block}
reaffirm the nonpositivity of the basis \eqref{basis-pa}.

Finally, note that the  basic states  keep their form in the
Majorana  representation too:
\begin{equation}
\label{basis-ma}
\Psi^{x_1\dots x_n}_{a_1\dots a_n}
= \gamma^{(1)}_{x_1,a_1}\ldots \gamma^{(1)}_{x_n,a_n} |0\rangle.
\end{equation}
Of course,  the Majorana fermions on the right side are in ascending
order.

\subsection{$O(N)$ structure of relative ground states}
\label{sub:gs}
In the previous section we have selected a nonpositive basis for the fermionic Hamiltonian \eqref{H}.
In addition, the Hamiltonian  connects any two basic elements belonging to an invariant
subspace \eqref{Va}.
Indeed, manipulating successively with the fermion hoppings $c_{x\pm 1,a}^+c_{x,a}$
and pair annihilations  $c_{x+ 1,a} c_{x,a}$, one can easily transfer any target basic state from the subspace
$ V^L_{a_1\dots a_m}$
to an $m$-particle \emph{trial} state. All particles there
 are gathered on the first site
of the chain \eqref{basis-order}:
\begin{equation}
\label{init}
\Psi_{a_1\dots a_m}=
\Psi_{a_1\dots a_m}^{1\,\dots\, 1}.
\end{equation}
The above wavefunction transforms as a rank-$m$ antisymmetric
tensor under the  rotations as already discussed \eqref{Yk}.

According to the Perron-Frobenius theorem,  the lowest energy state in
the invariant subspace  $ V^L_{a_1\dots a_m}$   (the \emph{relative} ground state)
  is nondegenerate.
Moreover, it can be expressed as a \emph{positive} superposition of all basic states
\eqref{basis}
 inside this subspace (denoted shortly by $\Phi_\alpha$):
\begin{equation}
\label{gs}
\Omega_{a_1\dots a_m}
=\sum_{\alpha=1}^{2^{N(L-1)}}\omega_\alpha \Phi_\alpha,
\qquad
\omega_\alpha >0.
\end{equation}
Since  the trial state \eqref{init} is a member of this family,
one can set $\Phi_1=\Psi_{a_1\dots a_m}$. Due to the rotational symmetry, the relative ground state
must be a part of a single $O(N)$ multiplet. Otherwise, it would split into mutually orthogonal
pieces,  belonging to nonequivalent multiplets. This fact would lead to a spontaneous symmetry breaking
 in the subspace $V^L_{a_1\dots a_m}$, which contradicts the above proven uniqueness condition.
Therefore, state $\Omega_{a_1\dots a_m}$ has
 the same $O(N)$ structure as  state $\Phi_1$ itself:
 Both wavefunctions belong to different but equivalent
 multiplets.

In particular, by removing the restriction on the indexes, one can set the lowest state
to be antisymmetric alike the trial one,
\begin{equation}
\label{gs-m}
\Omega_{\dots a_i\dots a_j \dots  }=-\Omega_{\dots a_j\dots a_i \dots}.
\end{equation}
Selecting another $m$-combination of the flavors, we arrive at a similar state
within the  subspace $V^L_{b_1\dots b_m}$. All such subspaces are equivalent as was
established in Sec.~\ref{sub:inv}, mapped to each other by the flavor exchanges \eqref{perm},  and produce
together a single degenerate $m$ sector  \eqref{W}.

Summarizing, we come to the conclusion that
\emph{the lowest energy  wavefunction  in the sector with $m$ odd flavors,
$\mathcal{V}^L_{m}$, is given by  a single $m$-th order antisymmetric $O(N)$ tensor
 described by the one-column Young tableau
of the same length:
\begin{equation}
\label{Y}
\yn_m
=\yn[1^m].
\end{equation}
The components provide the nondegenerate relative ground states  in the invariant
subspaces $V^L_{a_1\dots a_m}$.
}

Note that according to the representation theory of the orthogonal
group  \cite{Hamermesh},  the pair of multiplets, described by the Young diagrams
$\yn_m$ and  $ \yn_{N-m}$, are mutually conjugate and related by the  Levi-Civita symbol:
\begin{equation}
\label{conj}
\Omega'_{a_1\dots a_{N-m}}=\frac{1}{m!}\sum_{b_1,\dots,b_m}\epsilon_{a_1\dots a_{N-m}b_1\dots b_m}\Omega_{b_1\dots b_m}.
\end{equation}
As $SO(N)$ representations, they  are  equivalent and characterized by
the smallest number  among  $m$ and $N-m$.
Both multiplets are distinguished by the sign under improper rotations,
which maps tensor to  pseudotensor. One can mention this sign by the prime so that $O(N)$ representations
are characterized by the Young diagrams
$\yn_m$ and
\begin{equation}
\label{dual}
\yn_m'\sim\yn_{N-m}
 \quad
 \text{with} \quad
 m\le \sfrac12 N
\end{equation}
 provided that
for an even group rank,
$\yn_{N/2}\sim \yn'_{N/2}$.

For example, the empty diagram is a scalar (singlet) while the single $N$-length column,
 given by the Levi-Civita tensor,
is a pseudoscalar. So, according to our results, in the even-parity  sector, $\mathcal{V}^L_0$,
the lowest-level state is a scalar, whereas it  is a pseudoscalar
 in the odd-parity sector, $\mathcal{V}^L_{N}$.
Similarly, the lowest-level state in the sector with a single odd (even) flavor,
$\mathcal{V}^L_{1}$ ($\mathcal{V}^L_{N-1}$)
 is a vector (pseudovector).

\medskip

The relation among  the lowest energy levels  within the distinct invariant $m$ sectors
remains an open question. In particular, the \emph{total} ground state  may coincide with
a single antisymmetric multiplest $\yn_m$ for some $m$ or it can be an arbitrary combination
of them.
Below we  show that for the particular values of the coupling constants one can achieve the complete
degeneracy when the lowest energy levels  in all  sectors coincide. In that case, the relative ground states \eqref{gs}
from all  subspaces $V^L_{\sigma_1\dots \sigma_N}$ form the $2^N$-fold completely degenerate
total ground state.

\subsection{Decoupled Kitaev chains}
Consider the  chain \eqref{H}  without the four-fermion interactions where we set also $t_x=r_x$,
\begin{equation}
\begin{aligned}
H&=\sum_{a=1}^N H_a,
\\
H_a&=-\sum_{x=1}^{L-1} t_x\big( c^+_{x+1,a}c_{x,a}  +  c_{x+1,a}c_{x,a}\,+\text{H.c.}\big).
\end{aligned}
\end{equation}
The related Majorana system \eqref{HMaj} is reduced to the $N$ decoupled  Kitaev chains \cite{kitaev01}:
\begin{equation}
H_a=-\imath  \sum_{x=1}^{L-1}t_x\gamma^{(1)}_{x+1,a}\gamma^{(2)}_{x,a} .
\end{equation}

In each Hamiltonian, the two \emph{boundary} Majorana modes, $\gamma^{(1)}_{1,a}$ and $\gamma^{(2)}_{N,a}$,
are absent. They produce a single nonlocal fermion, the presence or absence of which does not affect  the
energy spectrum:
\begin{equation}
\big[H,c^\pm_{a}\big]=0,
\qquad
c_{a}^\pm=\frac{\gamma_{1,a}^{(1)}\mp\imath \gamma_{N,a}^{(2)} }{2}.
\end{equation}

One can choose  the boundary Majorana fermions, $\gamma_a=\gamma_{1,a}^{(1)}$,
to implement the mapping between different
$\sigma$ subspaces  \eqref{gamma-map}, \eqref{V++}.
In this case, they also intertwine  the Hamiltonian's action on these  subspaces.
Therefore, the spectrum in all subspaces   $V^L_{\sigma_1\dots \sigma_N}$  are identical.
 In particular, the ground state \emph{completely breaks} the $Z_2^{\times N}$ symmetry.

\section{Particle-hole symmetry in \boldmath{$O(N)$} fermion chain}
\label{sec:ph}
\subsection{Particle-hole transformation}

In this section, we define the particle-hole transformation and  study its properties.
For a single fermion, which we consider first, the particle-hole map may be described  as
a similarity transformation induced by the first Majorana fermion \eqref{gamma},
\begin{equation}
\label{ph}
\gamma^{(1)}c^\pm\gamma^{(1)}=c^\mp.
\end{equation}
A similar map, provided by the second fermion, produces an additional
sign:
\begin{equation}
\label{ph-}
\gamma^{(2)}c^\pm\gamma^{(2)}=-c^\mp.
\end{equation}
The above maps separate  two Majorana modes within a single
complex fermion: the first (second) map detects the parity of the
second (first) mode.

The transformations  \eqref{ph} and \eqref{ph-}  generate a $Z_2\times Z_2$
group and interchange   between the particle and hole,
$n\to 1-n$ with
$n=0,1$ meaning the fermion number.
Their composition 
 gives the parity operator  \eqref{na-major},
which alters the sign of the creation-annihilation operators:
$$
\sigma c^\pm \sigma =-c^\pm,
\qquad
\sigma=-\imath \gamma^{(1)}\gamma^{(2)}.
$$

Get back now to the chain model \eqref{H} and apply the last transformation
to all fermions located on the \emph{odd} nodes:
$c^\pm_{x,a}\to (-1)^x c^\pm_{x,a}$.
As  a result, it alters the signs
of the double-fermion couplings,
$$t_x\to -t_x,
\qquad
r_x\to -r_x,
$$
without
touching the other parts of the Hamiltonian. As a result, the positivity requirement on these
coefficients \eqref{pos} may be \emph{weakened} by setting the same sign for them:
\begin{equation}
\label{rt}
r_xt_x>0.
\end{equation}

Construct now a global \emph{particle-hole} map for the entire system in
a way suitable for our purposes.
Apply the transformations \eqref{ph} and \eqref{ph-}
to the even-site and odd-site fermions, respectively.
The resulted conjugation  is given by the following operator:
\begin{equation}
\label{p-h}
\hat\Gamma=
e^{-\imath\varphi}\prod_{x,a} \gamma_{x,a}^{(\lambda_x)},
\end{equation}
where  the function $\lambda_x$ separates the odd and even coordinates:
 $\lambda_\text{odd}=1$ and $\lambda_\text{even}=2$.
Although  the operator order in the product \eqref{p-h} is not relevant,
we set, for definiteness,  the ascending order  \eqref{order}.
As usual, the phase factor
\begin{equation}
\label{varphi}
\varphi=\pi\sfrac{(LN-1)LN}{4}
\end{equation}
is chosen to fulfill the involutivity condition:
\begin{equation}
\label{invol}
\hat\Gamma^2=1.
\end{equation}

Thus, the particle-hole operator $\hat\Gamma$ generates a $Z_2$ group.
 Obviously, it is unitary, which also ensures
the Hermiticity.
It (anti)commutes
with the Majorana fermion operators,
\begin{equation}
\label{ph-maj}
\begin{aligned}
\hat\Gamma\gamma^{(\lambda_x)}_{x,a}\hat\Gamma&=(-1)^{LN-1}\gamma^{(\lambda_x)}_{x,a},
\\
\hat\Gamma\gamma^{(\bar \lambda_x)}_{x,a}\hat\Gamma&=(-1)^{LN}\gamma^{(\bar\lambda_x)}_{x,a},
\end{aligned}
\end{equation}
as well as with the reflection operators, see Eq.~\eqref{chiral}:
\begin{equation}
\label{ph-sigma}
\hat\Gamma\hat\sigma_a\hat\Gamma=(-1)^{L}\hat\sigma_a.
\end{equation}
In addition, it  commutes with the proper
rotations \eqref{Lab} or \eqref{LabM}:
$$
\hat\Gamma\hat L_{ab}\hat\Gamma=\hat L_{ab}.
$$

The last two equations uncover the $O(N)$ structure of the $\hat\Gamma$.
It  is a \emph{scalar}  for  even-length chains
and  a \emph{pseudoscholar}  for odd lengths.

The global particle-hole map differs by a sign from its local counterparts
\eqref{ph} and \eqref{ph-}:
\begin{equation}
\label{ph-cpm}
\hat\Gamma c^\pm_{x,a}\hat\Gamma = (-1)^{LN-x} c^\mp_{x,a}.
\end{equation}
It converts the empty state into the completely
filled one with the prescribed fermion order \eqref{order}:
\begin{align}
\label{Gvac}
\hat \Gamma |0\rangle=e^{-\imath\varphi'}\prod_{x,a}c^+_{x,a}|0\rangle
=e^{-\imath\varphi'}|\overline{0}\rangle,
\\
\label{varphi'}
\varphi'=\varphi-\pi{\textstyle\frac N2 \left[\frac{L}{2}\right]}.
\end{align}
Here the barred vacuum
means  the empty-hole state.

A similar transform of  a general basic state \eqref{basis} produces an
additional sign factor, which may be calculated using
the relations \eqref{ph-cpm} and \eqref{Gvac}.
 In particular, a trial wavefunction \eqref{init}
converts  into the following state:
\begin{equation}
\label{phPsi}
\begin{aligned}
\hat \Gamma  \Psi_{a_1\dots a_m}
= e^{-\imath\varphi'}(-1)^{pL-m}\overline{\Psi}_{a_1\dots a_m},
\\[1mm]
p=(N-1)m+a_1+\dots+a_m.
\end{aligned}
\end{equation}
Here  again, the bar describes a state in terms of the holes rather
than particles. So, the state $\overline{\Psi}_{a_1\dots a_m}$
contains the ordered fermions, all  except those having the flavors
 $a_1,\dots, a_m$ and located on the first node:
\begin{equation}
\label{bPsi}
\overline{\Psi}_{a_1\dots a_m}=\prod_{(x,a)\ne (1,a_i)}c^+_{x,a}|0\rangle.
\end{equation}
It is a member of  the basis \eqref{basis}, or \eqref{basis-order}.

In general, any basic state $\Psi^{x_1\dots x_n}_{a_1\dots a_n}$
has its counterpart  $\overline{\Psi}^{x_1\dots x_n}_{a_1\dots a_n}$
with the holes instead of particles.
Clearly, there is no a $\hat \Gamma  $-invariant state,
so  the entire basis splits into $2^{NL-1}$
such pairs.

We remark that an equivalent particle-hole operator may be introduced
by applying the alternating local maps \eqref{ph} and \eqref{ph-}
in reverse order (see Ref.~\cite{Wei15} for the $N=1$ case):
\begin{equation}
\label{p-h'}
\hat\Gamma'=
e^{-\imath\varphi}\prod_{x,a} \gamma_{x,a}^{(\bar\lambda_x)}.
\end{equation}
Both matrices are related to each other through the total fermion parity
$\hat\sigma=\prod_{a=1}^N\hat\sigma_a$:
$$
\begin{aligned}
\hat\Gamma\hat\Gamma'&=\hat\Gamma'\hat\Gamma=\hat\sigma
\quad
\text{for even $NL$},
\\
\hat\Gamma\hat\Gamma'&=-\hat\Gamma'\hat\Gamma=
\imath\hat\sigma
\quad
\text{for odd $NL$}.
\end{aligned}
$$

\subsection{Particle-hole symmetric $O(N)$ chains}
\label{sub:phsym}

Remember that in Sec.~\ref{sub:gs} the $O(N)$ structure and degeneracy of the
lowest energy states of the fermionic model \eqref{H} is revealed.
Here we consider the behavior of these wavefunctions under the additional
$Z_2$  (particle-hole) symmetry.

The particular  choice, which distinguishes between the
even and odd sites \eqref{p-h},  implies the invariance of the
local Hamiltonian \eqref{Hyx}, modified by the replacement:
\begin{equation}
\label{modHyx}
\begin{aligned}
 P_{yx}^+P_{yx}&\to \frac12( P_{yx}^+P_{yx}+ P_{yx}P_{yx}^+),
 \\
K_{yx}^+K_{yx}&\to \frac12( K_{yx}^+K_{yx}+ K_{yx}K_{yx}^+).
\end{aligned}
\end{equation}
Indeed,  the relations \eqref{ph-maj} and
 Majorana fermion representations of the $K,P$ operators
 \eqref{Kmaj}, \eqref{Pmaj} imply
 \begin{equation}
\label{gkpg}
\hat\Gamma
\begin{pmatrix}
K_{x+1,x}\\
P_{x+1,x}
\end{pmatrix}
\hat\Gamma=
\begin{pmatrix}
K_{x+1,x}^+\\
P_{x+1,x}^+
\end{pmatrix}.
\end{equation}
Note that from the commutation relations \eqref{comKK}, it becomes clear that
the above modification \eqref{modHyx} results in an extra chemical potential,
$$
\delta V'=\sum_{x=1}^{L-1}(g_x+f_x)n_x+(g_x-f_x)n_{x+1},
$$
which may be absorbed by the potential.

From the other side,  the modified potential \eqref{V'} is not symmetric and
 undergoes the following  shift:
$$
\hat\Gamma V'(\dots, n_x,\dots )\hat\Gamma= V'(\dots, N-n_x,\dots).
$$

Consider now the potentials which remain \emph{invariant}   under
the particle-hole transformation:
$$
V'(n_1,\dots, n_L )=V'(N-n_1,\dots, N-n_L ).
$$
This happens, for example, when they depend on the products $(N-n_x)n_x$
as in the case of the $SU(N)$  Hubbard
potential. Therefore, the respective Hamiltonians are also preserved,
as can be inferred from the relations \eqref{Hoff}, \eqref{Hyx}, and
\eqref{gkpg}:
$$
[\hat \Gamma ,H]=0.
$$

The particle-hole structure of the relative ground states manifests
a \emph{parity effect} on the chain's size.

For \emph{even}-length chains, $L=2l$, the particle-hole  and reflection symmetries are
compatible according to the relation \eqref{ph-sigma}:
$$
[\hat \Gamma ,\hat\sigma_a]=0.
$$
Together they constitute a discrete group $Z_2^{\times (N+1)}$, which
preserves any individual $\sigma$ subspace.
Due to the uniqueness condition established in Sec.~\ref{sub:gs}, the relative ground state
also remains  \emph{invariant} under the particle-hole symmetry. To detect the corresponding quantum
number, we observe
that the basic states meet in pairs in the decomposition \eqref{gs}.  The pair members
are related by the particle-hole map such as, in particular, the two paired states,
 \eqref{init} and \eqref{bPsi}. The phase factor in the definition of $\hat \Gamma $
 \eqref{varphi'}
 is trivial,
 $\varphi'=\pi(Nl-1)Nl$,
 and Eq.~\eqref{phPsi}
 simplifies to the following one:
\begin{equation}
\label{phPsi'}
\hat \Gamma  \Psi_{a_1\dots a_m}
=(-1)^m \overline{\Psi}_{a_1\dots a_m}.
\end{equation}
Both states participate in the sum \eqref{gs} with positive coefficients, which
have to equal, giving rise to a combined state $\Psi_{a_1\dots a_m}+\overline{\Psi}_{a_1\dots a_m}$
with the particle-hole parity $(-1)^m$. Clearly, it
 coincides   with the eigenvalue of the relative ground state \eqref{gs}:
\begin{equation}
\label{phOmega}
\hat \Gamma  \Omega_{a_1\dots a_m}= (-1)^m\Omega_{a_1\dots a_m}.
\end{equation}

For \emph{odd}-length chains, $L=2l-1$, the  particle-hole transformation  anticommutes
with  reflections \eqref{ph-sigma},
\begin{equation}
\label{anti}
\{\hat \Gamma ,\hat\sigma_a\}=0.
\end{equation}
This fact leads to the additional twofold degeneracy of the
energy levels.  Indeed, the particle-hole transformation inverts
the parities  of all flavors. It matches                                                                                                                                                                                                                                                                                                                                                                                                                                                                                                                                                                        the Hamiltonian's
spectrum on  the two invariant $\sigma$ subspaces
\eqref{V}:
$$
\hat \Gamma V^L_{\sigma_1\dots\sigma_N}
=V^L_{-\sigma_1\dots-\sigma_N}.
$$
Therefore,  both subspaces have identical  spectra. As a consequence,
the two invariant sectors are degenerate \eqref{W},
$$
\hat \Gamma {\cal V}^L_m ={\cal V}^L_{N-m}.
$$
The exclusion is the sector with $m=N/2$ for \emph{even} values of the
group rank  $N$. The double degeneracy occurs
within the sector ${\cal V}^L_{N/2}$ containing an equal
number of flavors with odd and even parities.


\section{Further extensions and spin chains}

The result on the nondegeneracy and the multiplet structure of the relative ground state,
obtained in the previous section,  remain valid for more general class of $SO(N)$ invariant
fermionic chains. Recall that the local Hamiltonian \eqref{Hyx} is constructed from the
blocks \eqref{KP} using negative numbers to prevent any positive off-diagonal
matrix element in the selected basis \eqref{basis}.
In fact, more members can be added just keeping
this rule.

Note that the four-particle interactions in the original Hamiltonian \eqref{H}  are chosen to
\emph{preserve} the
number of fermions of each sort, $n_a$. The requirement simplifies the system but is not necessary.
Thus, the interaction   $K_{xy}P^+_{xy}$ with the conjugate  can also be included  in the local Hamiltonian
\eqref{Hyx}.
They are responsible for a particle decay into three particles  and the reverse process.

\label{sec:further}
\subsection{Long-range interactions}
So far, we have deal with the nearest-neighbor interaction in the open fermionic  chain.
The building blocks  of the Hamiltonian \eqref{KP}, which couple two distant sites, are not yet positive in the
 basis \eqref{basis-pa}. The simple replacement of the fermionic operators  with their bosonic counterparts
 given by the Pauli matrices \eqref{block} is not valid for a distant interaction any more. Instead,
  the Jordan-Wigner transformation \eqref{JW} imply a nonlocal coupling between distant sites,
  depending on the overall fermionic parity in all  intermediate positions. To avoid a sign problem,
  we redefine them  in the following way:
 \begin{equation}
 \label{Kxy'}
 \begin{aligned}
  K_{yx}=\sum_a\tau^+_{y,a}\tau^-_{x,a}
  =\sum_ac^+_{y,a}c_{x,a}\prod_{x<z<y} \sigma_{z,a},
\\
  P_{yx}=\sum_a\tau^-_{y,a}\tau^-_{x,a}
  =\sum_ac_{y,a}c_{x,a}\prod_{x<z<y}\sigma_{z,a}.
  \end{aligned}
\end{equation}
 Clearly, all  results
about the relative ground states and their multiplet structure,
established for the open chains, remain valid for an analog of the Hamiltonian \eqref{Hoff}
with the  \emph{long-range} interactions:
\begin{equation}
\label{Hlong}
\hat H_\text{lr}=\sum_{x<y} \left(H_{yx} + H_{yx}^+\right)+ V'(n_1,\dots,n_L).
\end{equation}
The interaction between two distant sites depends on \emph{positive} coupling constants,
as in the adjacent case  \eqref{Hyx}:
\begin{equation}
\label{Hyx'}
\begin{aligned}
H_{yx}  =&-t_{yx}  K_{yx} - r_{yx} P_{yx}
        - f_{yx} K_{yx}K_{yx}^+
        \\
        &- g_{yx}  P_{yx}^+P_{yx}-h_{yx} K_{yx}^2-e_{yx} P_{yx}^2.
\end{aligned}
\end{equation}

\subsection{Cyclic boundaries and translation invariance}
Let us restrict the distant interactions  to a single term  binding together
the first and
last sites.
As a result,  the nearest-neighboring   chain \eqref{Hoff} is supplemented by the \emph{cyclic} boundary term,
\begin{equation}
\label{Hb}
\hat H_\text{cyc}=\hat H+H_\text{b}, \qquad
H_\text{b}=H_{L1} + H_{L1}^+.
\end{equation}
As soon as this term is borrowed from
the long-range interacting model \eqref{Hyx'}, it fulfills
the sign rule.
It is easy to observe that the sign factors, depending on the intermediate
fermions, are provided by the total  parity operators  \eqref{Kxy'},
\begin{equation}
 \label{KL1}
  K_{L1}=-\sum_a c^+_{L,a}c_{1,a}\hat\sigma_a,
\quad
P_{L1}=-\sum_a c_{1,a}c_{L,a}\hat\sigma_a,
\end{equation}
which take constant values, $\sigma_a$, on the invariant subspaces $V^L_{\sigma_1\dots \sigma_N}$.
Thus, the boundary conditions depend on the individual $\sigma$ subspaces. Below we consider
only one of them.

Define  the elementary lattice translation:
\begin{equation}
\label{Tcpm}
\hat T c^\pm_{x,a} \hat T^{-1} = c^\pm_{x(\text{mod}\,L)+1,a}.
\end{equation}
Its eigenvalues $e^{ip}$ are given by the  lattice momentum values,
 $p=0,\sfrac{2\pi}{L},\dots,\sfrac{2\pi(L-1)}{L}$.
Evidently, it commutes with the $O(N)$ symmetry, including the parity operators:
\begin{equation}
\label{Tsig}
[\hat T,\hat \sigma_a]=0.
\end{equation}

Let us confine the Hamiltonian to the \emph{odd}-parity sector, ${\cal V}^L_N=V^L_{-\dots -}$,
 and consider the \emph{site-independent} coupling constants \eqref{pos}:
 $$
 r_x=r, \quad t_x= t, \quad f_x=f, \quad \text{etc.}
 $$
 Then the translation operator maps the local Hamiltonians  to each other,
 including
the boundary one \eqref{Hb}. This property ensures the \emph{translation invariance} of
the restricted system,
$$
\hat TH_{x\,x-1}\hat T^{-1}=H_{x+1\,x},
\qquad
[\hat T,\hat H_\text{cyc}]=0,
$$
with the  $x\pm1$ taken on modulo $L$ [see Eq.~\eqref{Tcpm}].

We affirm that
\emph{the relative ground state of the translation-invariant Hamiltonian  in the odd-parity sector, $\sigma_a=-1$,
is a pseudoscalar with zero momentum}:
\begin{equation}
\label{Tgs}
 \hat T \Omega_{12\dots N}=\Omega_{12\dots N}.
\end{equation}

One can use arguments similar to those in the proof of Eq.~\eqref{phOmega}.
Using the translations, circulate the trial state \eqref{init} to all nodes and take
the sum to get a translation-invariant state with zero momentum:
\begin{equation}
\label{initx}
\Psi=
\sum_{l=0}^{L-1}\hat T^l\Psi_{12\dots N},
\qquad
\hat T\Psi=\Psi.
\end{equation}
It is easy to observe that the above state takes part in the linear combination
\eqref{gs}. Due to the uniqueness condition, the both states, $\Omega_{12\dots N}$ and
$\Psi$,
have the same momentum quantum number, which proves the equation
\eqref{Tgs}.

Note that for the even-size chains with  odd Majorana modes per site
(which is not our case with the $2N$ modes), the commutator
in Eq.~\eqref{Tsig}
is replaced by the anticommutator.
As a result, the twofold degeneracy appears with the
supersymmetry behind it \cite{super}.
This resembles a similar behavior (without a supersymmetry)  in the case
of the particle-hole symmetry  and odd-size chains [see Sec.~\ref{sub:phsym},
Eq.~\eqref{anti}].

\subsection{Mixed-spin chains}
\label{sub:spin}

The six and more fermion exchange  terms may also contribute in the local interaction.
Among them there are the second- and  higher-order $SO(N)$ Heisenberg spin exchanges.

Consider the limiting case when all other terms are absent so that the Hamiltonian acquires the
following form:
\begin{equation}
\label{Hspin}
\hat H_\text{s} =     \sum_x \left(J^{(1)}_x \bm{L}_{x+1}\cdot \bm{L}_x -  J^{(2)}_x (\bm{L}_{x+1}\cdot \bm{L}_x)^2\right),
\end{equation}
where  higher powers, $k<N$, of the Heisenberg exchange
\eqref{spin-spin} may be involved too.
  In order to fulfill the required sign rule, they must be
with alternating couplings:
\begin{equation}
\label{Jk}
(-1)^{k-1}J^{(k)}_x
\qquad
\text{with}
\qquad J^{(k)}_x>0.
\end{equation}
The positivity condition may be
weakened (see, for example, the inequality  \eqref{J2}
for the second coupling below).

In fact, it is easy to check that the local spin-exchange
interaction \eqref{spin-spin}
is purely off-diagonal and
may be presented in the form,
\begin{equation}
\label{spin-spin'}
\bm{L}_y\cdot \bm{L}_x
=-\sum_{a\ne b}\left( c^+_{x,a}c_{y,a} c^+_{y,b} c_{x,b}+c^+_{x,a}c^+_{y,a} c_{y,b}c_{x,b}\right)
\end{equation}
with $y=x+1$.
This expression is built from the  double-fermion operators
with the same flavor, like $c^+_{x,a} c_{y,a}$.
As was derived already in Sec.~\ref{sec:basis},  they produce
non-negative matrices   in the current basis.
As a result, all matrix elements of   $\bm{L}_{x+1}\cdot \bm{L}_x$
are nonpositive.

Clearly, the Hamiltonian \eqref{Hspin} keeps unchanged  the local intrinsic spins
given  by the antisymmetric representations $\yn_{m_x}$
and  the total number of fermions per site:
$$
[\hat H_\text{s},n_x]=0.
$$
Thus,  it has a block diagonal form with $L^{N+1}$ parts,
according to all possible distributions
of the local
fermion numbers $n_x=m_x=0,\dots,N$ along the chain nodes. Moreover, the trivial representations, $m_x=0,N$,
appearing anywhere, cut the chain  into the two disjoint pieces.
We get in this way a set of the mixed-spin chains containing no more than  $L$ nodes.
Each site is endowed with an $O(N)$ antisymmetric multiplet (higher spin)  $\yn_m$
formed by  $n_x=m_x$ fermions provided that
\begin{equation}
\label{nontriv}
1\le m_x\le N-1.
\end{equation}

The particle-hole transformation $\hat\Gamma$ \eqref{ph} intertwines
 between  two "dual" chains composed
from the mutually conjugate representations with $m_x$ and $N-m_x$ fermions
per node,  ensuring the
equivalence of both Hamiltonians.

\medskip

Select now a \emph{single chain} from this family  and keep the notations
(for the length, invariant sectors, the lowest level  states there, etc.)
the same to avoid new entries.
Clearly,  the total amount of fermions and the total parity $\hat\sigma=(-1)^{\hat n}$
take \emph{constant} values therein:
$$
\begin{aligned}
n&=\sum_{a=1}^N n_a=\sum_{x=1}^L m_x,
\\
\sigma&=\prod_{a=1}^N \sigma_a=\prod_{x=1}^L(-1)^{m_x}.
\end{aligned}
$$
As a result, the reflection symmetry for spin chains is reduced
to the $Z^{\times (N-1)}$ group composed from $N-1$ independent parities
\cite{H15}.

Hence,  the allowed  invariant sectors
$\mathcal{V}^L_m$ \eqref{W} have to obey the \emph{parity rule}:
\begin{equation}
\label{rule}
(-1)^m=\sigma=(-1)^n.
\end{equation}
Of course, they are still $\binom{N}{m}$-fold degenerate, but
 their dimensions are essentially less comparing with those in the parent, fermionic chain.
Note that a single Majorana fermion $\gamma_a$ alters  the total parity value,
taking beyond the spin chain's space
of states. Therefore, the equivalence relation between two $\sigma$ subspaces
 \eqref{gamma-map} is not valid any more.
 Moreover,  their dimensions  differ  in general.

\begin{figure}[t]
\begin{center}
\includegraphics{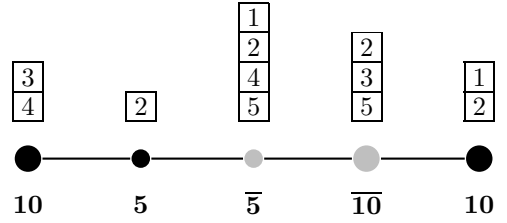}
\end{center}
\caption{\label{fig:Ochain} The five-site mixed-spin chain with $O(5)$
symmetry with  $x$th node  filled by the $m_x$ fermions.
The related antisymmetric multiplet
is characterized  by its dimension $\binom{5}{m}$ with or without the bar  listed below.
The bar (and gray color) marks a multiplet $m'=3,4$ dual to that with  $m=5-m'$,
A particular state \eqref{o5state} is shown
with the flavors inside the boxes atop.
}
\end{figure}

 Figure~\ref{fig:Ochain} illustrates a sample mixed-spin chain with $N=5$ flavors.
The (higher) spins on the nodes are members of the
antisymmetric multiplets $\yn_{m}$ with $m=2,1,3,4,2$ fermions per site.
Their dimensions are listed at the bottom.
For a conjugate pair $\yn_1,\yn'_1$ or $\yn_2,\yn'_2$
\eqref{dual}, they coincide so  the
representations  $\yn'_{1,2}$ are plotted in gray and marked with an overbar.

A sample state is displayed  in terms of the
Young tableau   at the top: a box means a
fermion with inscribed flavor \cite{Hamermesh}.
Using the comma to separate
the adjacent nodes apart, it may be written as follows:
\begin{equation}
\label{o5state}
\kett{34,2,1245,235,12}=
c^+_{1,3} c^+_{1,4}c^+_{2,2}
\dots
\dots
 c^+_{5,2} \kett{0}.
\end{equation}

\medskip
In the particular case when each node of the system \eqref{Hspin} is occupied by a single fermion,
$m_x=1$, we arrive at the $SO(N)$  invariant spin chain in the vector (defining)
representation, already considered in the current context \cite{harada14,H15}.
Following the
common parity rule \eqref{rule}, the total parity
must equal  the length's parity:
$$
(-1)^m=\sigma=(-1)^L.
$$

Rearranging fermions according to their positions as  done above \eqref{o5state}, we present the related
fermionic basis \eqref{basis} in a form more conventional for spin systems
(commas are omitted):
$$
|a_1\dots a_L\rangle =(-1)^{p_{a_1\dots a_L}} c^+_{1,a_1}\dots c^+_{L,a_L} |0\rangle.
$$
Here $p_{a_1\dots a_L}$ is a number of disordered pairs is the sequence, i.e., the
amount of the flavor pairs with $x<y$ but $a_x>a_y$.

The local spin exchange  \eqref{spin-spin'} produces on the above state:
$$
\bm{L}_{x+1} \cdot \bm{L}_x\, |ab\rangle=(-1)^{1-\delta_{ab}}|ba\rangle - \delta_{ab}\sum_c  |cc\rangle.
$$
Here only the spins at the neighboring $x$th and $(x+1)$th positions are mentioned, the others are not affected
and omitted.
As was already discussed, it contains solely negative elements out of the diagonal. Therefore, its square
 has positive matrix elements:
$$
 (\bm{L}_{x+1}\cdot \bm{L}_x)^2 |ab\rangle = |ab\rangle + (N-2) \delta_{ab}\sum_c  |cc\rangle
$$
 (the angular momentum exists merely for  $N\ge 2$).

The assembly of the common terms in both operators broadens the definition
area of the second coupling \eqref{Jk} beyond the positive region \cite{H15}:
\begin{equation}
\label{J2}
J_x^{(2)}>-\frac{J_x^{(1)}}{N-2}.
\end{equation}
It contains the integrable point,
\begin{equation}
\label{Jresh}
J^{(2)}=-\frac{N-4}{(N-2)^2}J^{(1)},
\end{equation}
at which the translational invariant system is solved exactly by the
Bethe ansatz \cite{resh}.

\medskip

One can spread out the results in Sec.~\ref{sub:gs} to the spin chain system.
In particular,
 \emph{the ground  state  in the sector  $\mathcal{V}^L_{m}$ is given by
 a single $m$-th order antisymmetric $O(N)$ tensor
 with the components $\Omega_{a_1\dots a_m}$, producing the unique relative ground states in
 the subspaces $V^L_{a_1\dots a_m}$.
}

The proof repeats the steps for the parent model
from Sec.~\ref{sec:lowest}.
The connectivity of the spin  Hamiltonian \eqref{Hspin}
in the nonpositive basis \eqref{basis-order} inside a restricted $\sigma$ subspace is easy to
establish using the representation \eqref{spin-spin}.
Due to the uniqueness and continuity, the multiplet type of the relative ground state remains
unchanged along the path connecting the Hamiltonians \eqref{H} and \eqref{Hspin}.
Alternatively, one can look for a more complex trial wavefunction  than the state \eqref{init},
which may go beyond the space of the spin chain states, see Refs.~\cite{H10,H15}.
For instance, one can set
\begin{equation}
\label{trial'}
\Psi_{a_1\dots a_m}=\sum_s \Psi_{a_1\dots a_m}^{x_{s_1}\dots x_{s_m}},
\end{equation}
where the sum is taken over the nontrivial permutations of a chosen coordinate set $x_1,\dots, x_m$.
It is easy to see that the above wavefunction is antisymmetric in the flavors.

\section{Conclusion}

We have studied the properties of the sectoral ground states (degeneracy,
multiplet structure) of the $O(N)$ invariant finite-size  chain of
interacting (Dirac or Majorana) fermions with  $N$ spins (flavors).

The system does not retain the number of particles with a given flavor $a$ but
keeps its parity, $\sigma_a$. The corresponding invariant subspaces
$V_{a_1\dots a_m}$  labeled by the values of odd flavors
(that is, the flavors with $\sigma_a=-1$), are related with each other
 by the Majorana modes
and thus have the same dimension.
In general, they differ in their spectrum.
Merely the  $\sigma$ subspaces with the same number $m$ of odd flavors
are completely degenerate and unified into a single
invariant $m$ sector.

For a wide range  of  coupling constants, we have  established in the current paper that
the lowest energy $O(N)$ multiplet  in any such $m$ sector is unique and
represented by an $m$th-order antisymmetric tensor, $\Omega_{a_1\dots a_m}$.
Owing  to the degeneracy, its components are the unique lowest energy states
(the relative ground states) in the subspace $V_{a_1\dots a_m}$.

Note that there is no  way to somehow relate the lowest levels of two distinct
sectors. In fact, the exact results in the two-site samples and   Kitaev chain
illustrate that there is not any ordering between them for common parameters.
The additional degeneracies  may happen at  special values of couplings,
 including the complete degeneracy
among all $N+1$ sectors.

Such degeneracies usually emerge in the presence of extra symmetries
as the particle-hole symmetry.
The impact on the spectrum depends on the parity of the chain's size.
  For the even-length
 chains, the particle-hole and  $O(N)$ symmetries commute.
 This endows the ground states $\Omega_{a_1\dots a_m}$ with
 an additional parity given by the   particle-hole eigenvalue,
 $(-1)^m$.
 For the odd length chains, the particle-hole map does not commute
 with improper rotations any more. Instead, it alters the values of
 all parities, leading
 to an additional twofold degeneracy between the
 dual subspaces $V_{a_1\dots a_m}$ and $V_{a'_1\dots a'_{N-m}}$
with differing flavor  sets $a_i$ and $a'_i$.

Remember that for the  $U(N)$ fermionic chain, there is no a particle
creation-annihilation process, so the fermion quantum numbers per flavor,
$n_a$, are good  and mark the invariant subspaces.
The subspaces with the same total number, $n=\sum_an_a$, are combined
into the sectors.
It is known that the lowest-energy states in each such $n$-sector
is described by a unique $m$th-order antisymmetric $U(N)$  multiplet
with $m=n\mod N$  \cite{H10}.

The above results remain valid
 in the presence of the long-range interactions
multiplied by a Jordan-Wigner string.
For the translation-invariant system, they are verified  merely in the odd-parity sector, $m=N$,
where the momentum quantum number of the ground state vanishes.

Finally, the aforementioned statements for the  fermionic models have been transferred as well
to the  $O(N)$ invariant, polynomial Heisenberg chains with alternating couplings and mixed spins.
This  system is obtained in the limit  when the fermion numbers (but not the flavors)
per each site persist. They set  ranks
of the antisymmetric multiplets, where the local spins live.
Note that now the  number of odd flavors ($m$) in each invariant subspace $V_{a_1\dots a_m}$
 is confined by the total number of fermions: both must have the same parity.
The obtained results extend those for the bilinear-biquadratic model
in the vector representation  \cite{H15}.

\begin{acknowledgments}
This work was   supported by the Armenian  Committee of Science Grant No.~18T-1C106.
The author is grateful to  anonymous referees for interesting and useful
remarks which led to the substantial improvement of  the current paper.
In particular, Sec.~\ref{sub:L=2}, the Appendix,
and a part of  Sec.~\ref{sub:spin} appeared due to them.
\end{acknowledgments}

\appendix

\section{Two-site $O(2)$ fermionic chain}
\label{sec:o2}

In the current Appendix we derive exactly the spectrum and quantum numbers of the particle-hole
invariant two-site $O(2)$ fermionic chain \eqref{H2site}, \eqref{o2KP}.
The obtained results have been described briefly  in Sec.~\ref{sub:L=2}.

The space of states decomposes into
the  invariant sectors ${\cal V}_m$
with $m=0,1,2$  \eqref{o2W}, \eqref{o2Va}.
Here we consider each sector separately.

\subsection{$m=0$ sector}
The sector ${\cal V}_0$ with even particle numbers is spanned by
the states
\begin{equation}
\label{bas++}
\kett{0},
\qquad
\kett{\up\dw,\up\dw},
\qquad
\kett{\up,\up},
\qquad
\kett{\dw,\dw},
\end{equation}
with a comma separating the nodes, see also the wavefunction \eqref{o5state}.
We use here the
conventional  basis where the fermion's position prevails over the spin. Three states
obey the non-positivity condition \eqref{Ax}, \eqref{basis} while the
fully occupied state,
$$
\kett{\up\dw,\up\dw}=c^+_{1,\up}c^+_{1,\dw}c^+_{2,\up}c^+_{2,\dw}\kett{0},
$$
needs a sign.

The particle-hole transformation \eqref{p-h} shuffles the states as follows
(see also Appendix~\ref{sub:pho3}):
\begin{equation}
\label{G++}
\hat\Gamma\kett{0}=
-\kett{\up\dw,\up\dw},
\qquad
\hat\Gamma\kett{\up,\up}=\kett{\dw,\dw}.
\end{equation}

Due to the Pauli exclusion principle, the fermion hopping
is banned here ($K=0$)  while the  annihilation of a fermion
pair is allowed  only on the following states:
\begin{equation}
\begin{gathered}
P\kett{\up\dw,\up\dw}=-\kett{\up,\up}-\kett{\dw,\dw},
\\
P\kett{\up,\up}=P\kett{\dw,\dw}=\kett{0}.
\end{gathered}
\end{equation}

The $O(2)$ symmetry splits the Hamiltonian into two diagonal
blocks.
The first block,
$$
H=\begin{pmatrix}
-2g & 2e & -\sqrt{2}r
\\
2 e & -2 g & \sqrt{2} r
\\
 -\sqrt{2} r & \sqrt{2} r &  -4 g
\end{pmatrix},
$$
 is spanned by the pure singlets:
\begin{equation}
\label{bas++'}
\kett{0},
\qquad
\kett{\up\dw,\up\dw},
\qquad
\sfrac{1}{\sqrt{2}}\left(\kett{\up,\up}+ \kett{\dw,\dw}\right).
\end{equation}

Note that the last one is specific for the orthogonal
groups. It forms a trace which contracts the spins of neighboring fermions
with  the invariant metrics, $\delta_{ab}$.
The traceless part of the symmetric tensor \eqref{psiab} makes up an
$O(2)$ doublet, one member of which,
\begin{equation}
\label{sym'1}
\sfrac{1}{\sqrt{2}}\left(\kett{\up,\up}-\kett{\dw,\dw}\right),
\end{equation}
forms the second block with $H=0$. [The second member  belongs
to the odd parity sector described below by Eq.~\eqref{sym'2}].
%

As a result, the energy eigenvalues are given by the following expressions,
\begin{equation}
\label{E++}
\begin{aligned}
& E_{0,1}=-e-3 g \mp \sqrt{(e-g)^2 + 4 r^2},
\\
& E_2=2 (e - g),
\qquad
E_3=0.
\end{aligned}
\end{equation}
Clearly, the lowest level, $E_0$ (top sign), is unique in the region with
positive parameters.

The relative ground state \eqref{gs} is
a singlet with the following coordinates  in the basis \eqref{bas++'}:
\[
\Omega=(\omega,-\omega,1),
\qquad
\omega=\sfrac{e - g + \sqrt{(e - g)^2 + 4 r^2}}{2 \sqrt{2} r}.
\]
Clearly, $\omega>0$, so  its coefficients in the basis
\eqref{basis} become positive in complete agreement with the most common
formula \eqref{gs}.

The particle-pole
parity is even, as is easy to see for the relations \eqref{G++},
\begin{equation}
\label{ph+}
\hat\Gamma\Omega=\Omega,
\end{equation}
 which agrees with the general
formula \eqref{phOmega}.

\subsection{$m=2$ sector}

The sector ${\cal V}_2$ with odd parities is spanned by two
fermions with opposite spins:
\begin{equation}
\label{bas--}
\kett{\up,\dw},
\qquad
\kett{\dw,\up},
\qquad
\kett{\up\dw,0},
\qquad
\kett{0,\up\dw}.
\end{equation}
Note that merely the second state has a wrong sign to make up
a nonpositive state \eqref{basis}.
The particle-hole inversion \eqref{p-h} produces
\begin{equation}
\label{G--}
\hat\Gamma\kett{\up,\dw}=
-\kett{\dw,\up},
\qquad
\hat\Gamma\kett{\up\dw,0}=\kett{0,\up\dw}.
\end{equation}

In contrast to the previous sector,
here the annihilation is forbidden, $P=0$, and the
particle motion conforms to the forward jumps,
\begin{equation}
\begin{aligned}
&K\kett{\up\dw,0}=\kett{\up,\dw}-\kett{\dw,\up},
\\
&K\kett{\up,\dw}=-K\kett{\dw,\up}=\kett{0,\up\dw},
\end{aligned}
\end{equation}
together with their backwards.

Make use of the $O(2)$ invariance again and split
the Hamiltonian into two independent parts. The first block
is built on the pseudoscalar sector,
\begin{equation}
\label{bas--'}
\sfrac{1}{\sqrt{2}}\left(\kett{\up,\dw}- \kett{\dw,\up}\right),
\qquad
\kett{\up\dw,0},
\qquad
\kett{0,\up\dw},
\end{equation}
with the following matrix,
$$
H=-\begin{pmatrix}
   4 f &  \sqrt{2} t &  \sqrt{2} t
  \\
   \sqrt{2} t &  2 f &  2h
 \\
   \sqrt{2} t   & 2h &   2 f
\end{pmatrix}.
$$

The symmetric combination,
\begin{equation}
\label{sym'2}
\sfrac{1}{\sqrt{2}}\left(\kett{\up,\dw}+\kett{\dw,\up}\right),
\end{equation}
is a zero-energy eigenstate.
Together with the
state
\eqref{sym'1}, it forms an $O(2)$ doublet described by the two-component
 tensor \eqref{psiab}.

Thus, the resulting energy levels  are,
\begin{equation}
\label{E--}
\begin{aligned}
& E_{0,1}=-h-3 f \mp \sqrt{(f-h)^2 + 4 t^2},
\\
& E_2=2 (h - f),
\qquad
E_3=0.
\end{aligned}
\end{equation}
Clearly, the lowest value, $E_0$, is unique in the region with
positive parameters.  The corresponding  state  is
a pseudo-singlet with the following coordinates in the basis \eqref{bas--'}:
\[
\Omega_{\up\dw}=(\omega,1,1),
\qquad
\omega=\sfrac{f-h+\sqrt{(f-h)^2+4t^2}}{\sqrt{2}t}.
\]
Since $\omega>0$, all coefficients in the basis \eqref{basis} are positive in agreement
with the general rule \eqref{gs}, as is easy to verity.
Its particle-pole
parity   is even,
$$
\hat\Gamma\Omega_{\up\dw}=\Omega_{\up\dw},
$$
 which follows from Eqs.~\eqref{G--}.
This fact agrees with the general formula \eqref{phOmega}.

\subsection{$m=1$ sector}
The sector ${\cal V}_1$  with mixed parities consists of the two $\sigma$ subspaces: $V_\up$
and $V_\dw$ \eqref{o2W}, spanned, correspondingly, by the first and second rows below:
\begin{align}
\label{basup}
&\kett{\up,0},
\qquad
\kett{0,\up},
\quad
-\kett{\dw,\up\dw},&
\quad
\kett{\up\dw,\dw},&
\\
\label{basdw}
&\kett{\dw,0},
\qquad
\kett{0,\dw},
\qquad
\kett{\up,\up\dw},&
-\kett{\up\dw,\up}.&
\end{align}
It is easy to see that every column forms a vector doublet, and the upper component
goes to the down one under the $\pi$ rotation.

The Hamiltonian \eqref{H2site} is given by  the same matrix in both  subspaces, $V_\up$ and $V_\dw$:
\begin{equation}
\label{Hup}
H=-\begin{pmatrix}
f + g & t & 0 & r
\\
 t & f +g & r & 0
 \\
 0 & r &f + g & t
 \\
  r & 0 & t &f +g
\end{pmatrix}.
\end{equation}
This fact reflects the degeneracy and can be also verified
using  the  action of the $K,P$
operators (and their Hermitian conjugates). The
nonvanishing processes on $V_\up$  are provided by
$$
\begin{aligned}
K\kett{\up,0}&=\kett{0,\up},
\quad
&K\kett{\dw,\up\dw}&= -\kett{0,\up},
\\
P\kett{\dw,\up\dw}&= -\kett{0,\up},
\quad
&P\kett{\up\dw,\dw}&= \kett{\up,0},
\end{aligned}
$$
together with their backwards.

The corresponding energy levels are pretty simple:
\begin{equation}
\label{E+-}
E_{0,1,2,3}=-f-g\mp r\mp t.
\end{equation}
Obviously, the lowest level, $E_0$ (with minus signs),
is unique in the subspace $V_\up$ ($V_\dw$).

The relative ground states in both subspaces have the same expansion,
\begin{equation}
\label{Oupdw}
\Omega_\up=\Omega_\dw=(1,1,1,1).
\end{equation}
Clearly, both components  form a vector doublet, $\Omega_a$.
The eight basic states  \eqref{basup}, \eqref{basdw}   match the
nonpositivity condition \eqref{basis}, and the unit coefficients
justify the most common decomposition formula \eqref{gs}.

 Note that simple expressions for the spectrum and eigenstates
 are the consequence of the lattice reflection and
  particle-hole symmetries of the Hamiltonian \eqref{H2site}.
 Both of them shuffle the basic states \eqref{basup}, \eqref{basdw}.
 The lowest-energy state is even under the lattice reflection
and is odd under the particle-hole transformation:
 $$
 \hat\Gamma\Omega_a=-\Omega_a.
 $$
  The last fact is in agreement with the general rule
  \eqref{phOmega} and
 can be verified independently applying
$$
\begin{aligned}
\hat\Gamma\kett{\up,0}&=\kett{\dw,\up\dw},
\quad
&\hat\Gamma\kett{0,\up}&=-\kett{\up\dw,\dw},
\\
\hat\Gamma\kett{\dw,0}&=-\kett{\up,\up\dw},
\quad
&\hat\Gamma\kett{0,\dw}&=\kett{\up\dw,\up}
\end{aligned}
$$
and taking into account the involutivity property \eqref{invol}.

\section{Two-site $O(3)$ fermionic chain}
\label{sec:o3}

Here we derive exactly the spectrum and quantum numbers of the particle-hole
invariant two-site $O(3)$ fermionic chain \eqref{H2site}, \eqref{o2KP}.
The obtained results have been discussed briefly  in Sec.~\ref{sub:L=2}.

The space of states decomposes into
the  invariant sectors ${\cal V}_m$
with $m=0,1,2,3$ \eqref{o3W}, \eqref{o3Va}. Below we treat each sector separately.

\subsection{$m=0$ sector}
The sector with even parities, ${\cal V}_0$, is formed by the following
eight states (no  summation),
\begin{equation}
\label{bas+++}
\kett{0}, \quad
\kett{a,a},
\quad
-\kett{ab,ab},
\quad
-\kett{123,123},
\end{equation}
where $a,b=1,2,3$ and $a<b$.  We continue with
the Dirac notations from the previous section with
the comma separating the nodes.
The minus signs has been set to fit the nonpositivity condition \eqref{basis}.

 Due to the Pauli exclusion
principle, a particle hopping is forbidden here ($K=0$) while
the pair annihilation is subjected to the rule,
$$
\begin{gathered}
P\kett{a,a}=\kett{0},
\quad
P\kett{ab,ab}=-\kett{a,a}-\kett{b,b},
\\
P\kett{123,123}=\sum_{a<b}\kett{ab,ab}.
\end{gathered}
$$

The Hamiltonian is split into two blocks, each involving  equivalent
 multiplets.
The first block is made from the scalars (singlets), which we
endow with the following basis,
\begin{equation}
\label{bas-sing}
\kett{0}, \quad
\sum_a\frac{\kett{a,a}}{\sqrt{3}},
\quad
-\sum_{a<b}\frac{\kett{ab,ab}}{\sqrt{3}},
\quad
-\kett{123,123}.
\end{equation}

Inserting  the structure of the $K,P$ operators into the Hamiltonian \eqref{H2site},
we obtain:
\begin{equation}
\label{H+++}
H=-\begin{pmatrix}
3g & \sqrt{3}r  & 2\sqrt{3}e & 0
\\
 \sqrt{3}r & 7g & 2r & 2\sqrt{3}e
 \\
  2 \sqrt{3} e & 2 r & 7 g & \sqrt{3}r
 \\
0 & 2\sqrt{3}e & \sqrt{3}r & 3g
\end{pmatrix}.
\end{equation}
The energy levels are,
\begin{equation}
\label{E+++1}
\begin{aligned}
E_{0,1}=-5 g - r \mp 2 \sqrt{3 e^2 + g^2 + 3 e r + g r + r^2},
\\
E_{2,3}=-5 g + r \mp 2 \sqrt{3 e^2 + g^2 - 3 e r -  g r + r^2}.
\end{aligned}
\end{equation}
Among them, the level $E_0$ (with -- sign) is the lowest one, not
only among singlets but also in the whole sector as we will
see later. The related state has positive  coordinates
in the singlet basis:
\[
\begin{aligned}
\Omega&=(1,\omega,\omega,1),
\\
\omega&=\sfrac{2 g+r+2 \sqrt{3 e^2+g^2+3 e r+r (g+r)}}{\sqrt{3} (2 e+r)}>0.
\end{aligned}
\]
Then the coordinates in the current $\sigma$ subspace
 basis \eqref{bas+++} are also positive, as is easy to see,
in complete agreement with the expansion formula for the relative
ground state \eqref{gs}.

The second block  of the Hamiltonian is based on the diagonal
parts of the two equivalent $O(3)$ quintets,  described by the
traceless symmetric tensor \eqref{psiab},
\begin{equation}
\label{sym}
\psi_{ab}^{(1)}\ket{a,b},
\qquad
\frac14\psi^{(2)}_{ef}\epsilon_{eab}\epsilon_{fcd}\ket{ab,cd},
\end{equation}
where the summation is supposed over the repeating indexes.
Such states are characterized  by the Young tableau
\ytableausetup{centertableaux,boxsize=1.2em}
$\begin{ytableau}a & b \end{ytableau}$ \cite{Hamermesh}.

\smallskip

 We select the following diagonal tensors:
 \begin{equation}
 \label{diag-psi}
 \psi_{ab}=\psi_{aa}\delta_{ab}
 \quad
 \text{with}
 \quad
 \begin{matrix}
 1)\, \psi_{aa}=
 (1,-1,0),
 \\
  2) \,\psi_{aa}=(1,1,-2).
\end{matrix}
 \end{equation}
They define orthogonal states (normalization  is not essential here),
\begin{equation}
\label{bas-diag}
\begin{aligned}
&\kett{1,1}-\kett{2,2},
\quad
&&\kett{1,1}+\kett{2,2}-2\kett{3,3},
\\
&\kett{23,23}-\kett{13,13},
&&\kett{23,23}+\kett{13,13}-2\kett{12,12}.
\end{aligned}
\end{equation}

The Hamiltonian mixes together the states within the same
column,  so that
 is spit once more into two identical blocks,  each
 given by
$$
H=-\begin{pmatrix}
 g & r \\
r & g
\end{pmatrix}.
$$
Thus, the corresponding energy levels are doubly degenerate:
\begin{equation}
\label{E+++2}
E_{4,5}=-g - r,
\quad
E_{6,7}=-g + r.
\end{equation}
Each level is filled by the two diagonal states within the same symmetric multiplet.
Comparing the above  values  with the singlet  levels \eqref{E+++1}, we make sure
that the energy $E_0$ is indeed the lowest one in the current sector.

\subsection{$m=3$ sector}

The sector with odd parities, ${\cal V}_3$, is formed by the
three particles with different flavors. We write
them  in cyclic order for the latter convenience,
\begin{equation}
\label{bas---}
\begin{aligned}
\kett{123,0},
\quad
\kett{12,3},
\quad
\kett{23,1},
\quad
\kett{31,2},
\\
\kett{0,123},
\quad
\kett{1,23},
\quad
\kett{2,31},
\quad
\kett{3,12}.
\end{aligned}
\end{equation}
In this form, by the way, they all belong to the
nonpositive basis \eqref{basis}, as is easy to verify.

The  Pauli exclusion
principle  bans the creation and annihilation processes
($P=0$).  The jumps from left to right are driven by
\begin{equation}
\label{K123}
K\kett{123,0}=\kett{23,1}+\kett{31,2}+\kett{12,3},
\end{equation}
and six more rules given  by
\begin{equation}
\label{K12}
K\kett{12,3}=\kett{1,23}+\kett{2,31},
\quad
K\kett{1,23}=\kett{0,123}
\end{equation}
and their cyclic permutations. Evidently, the backward jumps
are driven by the operator $K^+$.

Here again the Hamiltonian \eqref{H2site} decomposes into diagonal blocks,
composed from equivalent $O(3)$ multiplets.
One such block is made up of the pseudoscalars (pseudo-singlets), which
we endow with the following basis:
\begin{equation}
\label{bas-psin}
\begin{aligned}
\kett{123,0}, \quad
\frac{\epsilon_{abc}}{2\sqrt{3}}\kett{ab,c},
\quad
\frac{\epsilon_{abc}}{2\sqrt{3}}\kett{a,bc},
\quad
\kett{0,123},
\end{aligned}
\end{equation}
with the summation over the Levi-Civita indexes.
Using the  hoppings, described above,
we derive the restriction of the Hamiltonian there:
\begin{equation}
\label{H---}
H=-\begin{pmatrix}
 3f &  \sqrt{3}t  &  2\sqrt{3}h & 0
\\
  \sqrt{3}t &  7f &  2t &  2\sqrt{3}h
 \\
   2 \sqrt{3} h &  2 t &  7 f &  \sqrt{3}t
 \\
0 &  2\sqrt{3}h &  \sqrt{3}t &  3f
\end{pmatrix}.
\end{equation}
The energy levels can be easily calculated:
\begin{equation}
\label{E---1}
\begin{aligned}
E_{0,1}=-5 f -t \mp 2 \sqrt{3 h^2 + f^2 + 3 ht + ft + t^2},
\\
E_{2,3}=-5 f + t \mp 2 \sqrt{3 h^2 + f^2 - 3 ht -  ft + t^2}.
\end{aligned}
\end{equation}

The second block of the Hamiltonian is formed on the
space spanned by the four states,
\begin{equation}
\label{bas-mix}
\begin{aligned}
&\kett{23,1}-\kett{31,2},
\quad
&\kett{23,1}+\kett{31,2}-2\kett{12,3},
\\
&\kett{1,23}-\kett{2,31},
&\kett{1,23}+\kett{2,31}-2\kett{3,12}.
\end{aligned}
\end{equation}

In fact, the above states  belong to the two equivalent
  $O(3)$ quintets based on the three-component  tensor with mixed symmetry \eqref{psiabc},
\begin{align}
\label{mixed}
\frac12\psi^{(1)}_{abc}\ket{ab,c},
\qquad
\frac12\psi^{(2)}_{abc}\ket{c,ab},
\end{align}
where the sum is taken over the repeating indexes.
In the representation theory terms, the states in
both multiplets are described by the
Young tableau
\ytableausetup{centertableaux,boxsize=1.2em}
$\begin{ytableau}a & c\\ b \end{ytableau}$.

\smallskip

It is a pseudo-analog to the $\psi_{ab}$ tensor (or its dual in the representation theory
terminology \cite{Hamermesh}). In particular, the above  states, like the  wavefunctions
\eqref{bas-diag},
are based on the diagonal tensors \eqref{diag-psi},
which produce mixed ones by
 $\psi_{abc}=\epsilon_{abc}\psi_{cc}$ (no sum).

Again, the Hamiltonian entangles only the  column states in the
table \eqref{bas-mix}. In each column, it acquires the  form
$$
H=\begin{pmatrix}
-f & t \\
t & -f
\end{pmatrix}.
$$
Thus, the energy levels are doubly degenerate:
\begin{equation}
\label{E---2}
E_{4,5}=-f - t ,
\quad
E_{6,7}=-f +t .
\end{equation}
Each level is filled by two states from a  quintet with mixed
symmetry being dual to the symmetric quintet.
Comparing with the energy levels \eqref{E---1},
we conclude that the lowest one is provided by the energy $E_0$.

The related relative ground state is  derived easily. Of course, it is a
non-degenerate
pseudo-singlet and has the positive
coefficients in the basis \eqref{bas-psin},
\[
\begin{aligned}
\Omega_{123}&=(1,\omega,\omega,1),
\\
\omega&=\sfrac{2 f+t+2 \sqrt{f^2+3 h^2+f t+3 h t+t^2}}{\sqrt{3} (2 h+t)}>0.
\end{aligned}
\]
It is easy to see that its coefficients  in the total basis \eqref{bas---}
are also positive. This agrees with the established law
\eqref{gs}.

\subsection{$m=1$ sector}
The sector with a single flavor with an odd particle number,
${\cal V}_1$, contains three equivalent $\sigma$ subspaces,
$V_a$ with $a=1,2,3$ \eqref{o3W}, \eqref{o3Va}.

First, consider the subspace $V_1$, where
the flavor $a=1$ is odd, and take the
following  basis there:
\begin{equation}
\label{bas++-}
\begin{gathered}
\kett{1,0},
\quad
\kett{0,1},
\quad
\kett{1b,b},
\quad
-\kett{b,1b},
\\
-\kett{123,23},
\quad
-\kett{23,123}
\end{gathered}
\end{equation}
with $b=2,3$ and without summation.
The signs are set in order to match the common rule \eqref{basis}.

The direct (left-to-right) moves are described by the hoppings:
$$
\begin{gathered}
K\kett{1,0}=\kett{0,1},
\quad
K\kett{1a,a}=-\kett{a,1a},
\\
K\kett{123,23}=\kett{23,123}.
\end{gathered}
$$
Similarly, the pair annihilations are managed by the rules:
$$
\begin{gathered}
P\kett{1a,a}=\kett{1,0},
\quad
P\kett{123,23}=-\kett{12,2}-\kett{13,3},
\\
P\kett{a,1a}=-\kett{0,1},
\quad
P\kett{23,123}=\kett{3,13}+\kett{2,12}.
\end{gathered}
$$

Note that the first two states in the basis \eqref{bas++-},
represent the first component of a vector. In fact, the last two
also represent the same. For example,
\begin{equation}
\label{bas-vec'}
\kett{123,23}=\frac12\sum_{a,b}\kett{1ab,ab}.
\end{equation}
Taking the sum over $b$ in Eqs.~\eqref{bas++-}, we get two
more vectors. We have extracted in this way the six-dimensional subspace
spanned  by the
first components of vectors,
\begin{equation}
\label{bas-vec}
\begin{gathered}
\kett{1,0},
\quad
\kett{0,1},
\quad
\sum_b\frac{\kett{1b,b}}{\sqrt{3}},
\quad
-\sum_b\frac{\kett{b,1b}}{\sqrt{3}},
\\
-\frac12\sum_{b,c}\kett{1bc,bc},
\quad
-\frac12\sum_{b,c}\kett{bc,1bc},
\end{gathered}
\end{equation}

Clearly, the Hamiltonian \eqref{H2site} preserves that subspace. It has the following matrix
representation  therein:
$$
H=-\begin{pmatrix}
u & t & \sqrt{2} r & 0 & 2 e & 0 \\
 t & u & 0 & \sqrt{2} r & 0 & 2 e \\
 \sqrt{2} r & 0 & u+2 g & t & \sqrt{2} r & 0 \\
 0 & \sqrt{2} r & t & u+2g & 0 & \sqrt{2} r \\
 2 e & 0 & \sqrt{2} r & 0 & u & t \\
 0 & 2 e & 0 & \sqrt{2} r & t &u
 \end{pmatrix}
 $$
 with the substitution  $u=f+2g$.
 The corresponding six energy levels are
\begin{equation}
\label{E++-1}
\begin{aligned}
E_{0,1,2,3}&=-3 g -e-f\mp t\mp\sqrt{(e-g)^2+4 r^2},
\\
E_{4,5}&=2 (e-g)-f\mp t.
\end{aligned}
\end{equation}

The second block of the Hamiltonian is formed on
the remaining two states, which are associated with the
two quintets with mixed symmetry \eqref{mixed},
\begin{equation}
\label{mix1}
\kett{12,2}-\kett{13,3},
\qquad
\kett{2,12}-\kett{3,13}.
\end{equation}
In more detail, the above states correspond to the $\psi_{abc}$
 \eqref{psiabc}
which is obtained from
the tensor
\begin{equation}
\label{psi23}
\psi_{ab}=\delta_{a2}\delta_{b3}+\delta_{a3}\delta_{b2}.
%
\end{equation}
The Hamiltonian on this subspace acquires the matrix form,
$$
H=\begin{pmatrix}
 -f & t\\
 t & -f
\end{pmatrix}
$$
with the energy spectrum given by,
\begin{equation}
\label{E++-2}
E_{6,7}=-f\pm t.
\end{equation}
We conclude that the minimal level corresponds to $E_0$  in the vector
spectrum \eqref{E++-1}.

The  basis \eqref{bas-vec} trivially
spreads to any of three subspaces $V_a$ by
replacing the first flavor
with  the $a$th. Let us denote the corresponding
basis by $\Psi_{ai}$ and keep the disposition, so
 $\Psi_{a1}=\kett{a,0}$,   $\Psi_{a2}=\kett{0,a}$,
etc.

Evidently, the Hamiltonian's matrix
is the same for all $V_a$.
The lowest-energy states of the current sector
are combined into a single vector
triplet, which may be derived explicitly,
 \begin{equation}
 \label{Oa}
\begin{aligned}
\Omega_a=\sum_{i=1}^6\omega_i\Psi_{ai},
\qquad
\omega_i=(1,1,\omega,\omega,1,1)
\\
\text{with}\quad \omega=\sfrac{g-e+\sqrt{(g-e)^2+4 r^2}}{\sqrt{2} r}>0.
\end{aligned}
\end{equation}
The coefficients in the $\sigma$-subspace  basis \eqref{bas++-}
are positive  too
 [compare with Eq.~\eqref{gs} for the general case].

The counterparts of \eqref{mix1}
in the subspaces $V_2$ and $V_3$
are obtained by cyclic permutations of the
flavors. They are all provided by the off-diagonal components of the symmetric tensor
$\psi_{ab}$ [see Eq.~\eqref{psi23}].
The  diagonal components \eqref{mixed}
 have already generated  the states  \eqref{bas-mix}
 of the same quintets in the odd sector ${\cal V}_3$.

\subsection{$m=2$ sector}
Consider now the sector ${\cal V}_2$ having two  flavors  with  odd particle numbers.
Alike the previous sector, it is also degenerate and contains three $\sigma$ subspaces,
$V_{ab}$, labeled by
the values of odd flavors \eqref{o3W}.

For definiteness, consider the  subspace $V_{23}$ and choose
the following  basis there:
\begin{equation}
\label{bas+--}
\begin{aligned}
\kett{23,0},&
\quad
\kett{123,1},&
-\kett{12,13},&
\qquad
\kett{2,3},&
\\
\kett{0,23},&
\quad
\kett{1,123},&
\quad
\kett{13,12},&
\quad
-\kett{3,2}.&
\end{aligned}
\end{equation}
Here the minus sign makes the basis  nonpositive.

The direct moves are provided  by the
jumps,
$$
\begin{aligned}
 K\kett{123,1}&=\kett{13,12}-\kett{12,13},
\\
K\kett{23,0}&=\kett{2,3}-\kett{3,2},
\\
K\kett{1a,1b}&=\kett{1,1ba},
\quad
K\kett{a,b}=\epsilon_{1ab}\kett{0,ab},
\end{aligned}
$$
 where $a,b=2,3$.
The following annihilations are also allowed:
$$
\begin{aligned}
& P\kett{123,1}=\kett{23,0},
\qquad
P\kett{1,123}=\kett{0,23},
\\
&P\kett{1a,1b}=-\kett{a,b}.
\end{aligned}
$$
The backward  processes are implemented
 by the Hermitian conjugate operators.

Let us extract the six-dimensional subspace spanned by the first
components of pseudo-vectors (equivalently, by the second and third components
of antisymmetric tensors). Define the basis with Levi-Civita symbol
making apparent their  structure,
\begin{equation}
\label{bas-pvec}
\begin{aligned}
&\frac{\epsilon_{1ab}}{2}\kett{ab,0},
\quad
&&\frac{\epsilon_{1ab}}{2}\kett{1ab,1},
&&-\frac{\epsilon_{1ab}}{\sqrt{2}}\kett{1a,1b},
\\
&\frac{\epsilon_{1ab}}{\sqrt{2}}\kett{a,b},
\quad
&&\frac{\epsilon_{1ab}}{2}\kett{0,ab},
\quad
&&\frac{\epsilon_{1ab}}{2}\kett{1,1ab}
\end{aligned}
\end{equation}
with the sum taken over the flavors $a,b$.

The Hamiltonian \eqref{H2site} has the following structure
therein:
 $$
H=-\begin{pmatrix}
u & r & \sqrt{2} t & 0 & 2 h & 0 \\
 r & u & 0 & \sqrt{2} t & 0 & 2 h \\
 \sqrt{2} t & 0 & u+2f & r & \sqrt{2} t & 0 \\
 0 & \sqrt{2} t & r & u+2f & 0 & \sqrt{2} t \\
 2 h & 0 & \sqrt{2} t & 0 & u & r \\
 0 & 2 h & 0 & \sqrt{2} t & r & u
 \end{pmatrix}
 $$
with the substitution $u=2f+g$. It has the following
eigenvalues:
\begin{equation}
\label{E+--1}
\begin{aligned}
E_{0,1,2,3}&=  -3 f-h-g\mp r\mp \sqrt{(f-h)^2+4 t^2},
\\
E_{4,5}&=2 (h-f)-g\mp r.
\end{aligned}
\end{equation}

The second block of the Hamiltonian is formed on
the states belonging to the two equivalent  symmetric
$O(3)$ quintets \eqref{sym}.
They correspond to the choice \eqref{psi23} for
 the symmetric tensor
\eqref{psiab}:
\begin{equation}
\label{sym1}
\kett{2,3}+\kett{3,2},
\qquad
\kett{12,13}+\kett{13,12}.
\end{equation}
Remember that the diagonal components of these quintets \eqref{bas-diag}
have already
taken part
in the even sector ${\cal V}_0$. Thus, the energy levels (excluding the degeneracy)
are inherited therefrom \eqref{E+++2}:
\begin{equation}
\label{E+--2}
E_{6,7}=-g\pm r.
\end{equation}
Of course, they may be also obtained from the restriction of the Hamiltonian
to the subspace spanned by the states \eqref{sym1}:
$$
H=\begin{pmatrix}
 -g & r\\
 r & -g
\end{pmatrix}.
$$

Thus,  the minimal energy level in the current sector
 belongs
to the pseudo-vector's spectrum and has the value $E_0$
\eqref{E+--1}.

The pseudo-vector basis \eqref{bas-pvec} may be
expanded to the whole sector replacing of the first  flavor
by  the $c$th one. Keeping the disposition, denote the corresponding
basis by $\Psi'_{ci}$.
Due to the degeneracy, the Hamiltonian is described by
the same matrix in all $V_{ab}$.

The lowest-energy states of the current sector
are gathered into the following single pseudo-vector
(two-component antisymmetric tensor)
triplet:
 \begin{equation}
 \label{Oab}
\begin{aligned}
\Omega_{ab}=\sum_{c,i}\omega_i\epsilon_{abc}\Psi'_{ci},
\qquad
\omega_i=(1,1,\omega,\omega,1,1)
\\
\text{with}\quad \omega=\sfrac{f-h+\sqrt{(f-h)^2+4 t^2}}{\sqrt{2} t}>0.
\end{aligned}
\end{equation}

The symmetric states  \eqref{sym1}
in the subspaces $V_{12}$ and $V_{13}$
are obtained by cyclic permutations of the
flavors. They are provided by the $\psi_{12}$
and $\psi_{13}$  components of the symmetric tensor
\eqref{psi23}.
Recall that the  diagonal parts  of the symmetric quintets \eqref{sym}
 have already participated in the even-parity sector ${\cal V}_0$.

\subsection{Particle-hole parity}
\label{sub:pho3}

Here we address  the properties of the particle-hole inversion.
Recall that the latter maps the empty state to the completely filled
one  \eqref{Gvac}:
\begin{equation}
\label{Gvac3}
\hat\Gamma \kett{0}=
|\overline{0}\rangle=-\kett{123,123}.
\end{equation}
The minus sign has arisen  due to the difference in
fermion disposition in the last two states.
Together with the relation \eqref{ph-cpm}, it determines
 the action of the $\hat\Gamma$ on an arbitrary  state.
Indeed, it can be verified that a common basic wavefunction
with $m$ fermions in the ascending order \eqref{basis-order}, \eqref{order}
obeys the same relation as that for the trial state \eqref{phPsi}. For
the $L=\emph{even}$ chains, it reduces to
\begin{equation}
\label{ph-basis}
\begin{aligned}
\hat \Gamma  \Psi_{a_1\dots a_m}^{x_1\dots x_m}
=(-1)^{m}\overline{\Psi}_{a_1\dots a_m}^{x_1\dots x_m}.
\end{aligned}
\end{equation}
Remember that a bared state lists the holes but not the particles \eqref{bPsi}.

According to the results in Sec.~\ref{sub:phsym}, the inverted
wavefunction, $\overline{\Psi}$,
is also a member of  the nonpositive basis \eqref{basis-order}. Moreover,
both states \eqref{ph-basis} belong to the same $\sigma$ subspace.

The above properties hold, in particular, for the basis in all sectors
${\cal V}_m$ with $m=0,1,2,3$, given, respectively by Eqs.~\eqref{bas+++}, \eqref{bas---}, \eqref{bas++-},
\eqref{bas+--}. The corresponding lowest-energy states
are positive superpositions of the
basic wavefunctions. Together with the relation \eqref{ph-basis}, this fact
set their particle-hole quantum numbers to $(-1)^m$.

Consider, for instance, the sector ${\cal V}_1$ (${\cal V}_2$),
where  the relative ground state is formed by the six vectors $\Psi_{a i}$ \eqref{bas-vec}
[pseudo-vectors $\Psi'_{ai}$ \eqref{bas-pvec}] numbered by index
$i$. Then the particle-hole inversion \eqref{ph-basis}
results  on them,
$$
\hat\Gamma \Psi_{ai}=-\overline\Psi_{ai}=- \Psi_{a\bar i}
\qquad
\big(\hat\Gamma \Psi'_{ai}=\overline\Psi'_{ai}= \Psi'_{a\bar i}\big),
$$
\\[0mm]
with $\bar i=7-i$. Applying the above map to the lowest-level
states \eqref{Oa} or \eqref{Oab}, we obtain
$$
\hat\Gamma\Omega_a = -\Omega_a
\qquad
\big(\hat\Gamma\Omega_{ab} = \Omega_{ab}\big),
$$
in complete agreement with the formula \eqref{phOmega}
established for the common even-length chains.


\begin{thebibliography}{99}


\bibitem{M55}
W. Marshall,
\emph{Antiferromagnetism}, \\
\href{http://dx.doi.org/10.1098/rspa.1955.0200}{Proc. R. Soc. A {\bf 232}  (1955) 48}.

\bibitem{LSM61}
E. H. Lieb, T. D. Schultz, and D. C. Mattis, 
\emph{Two soluble  models of an antiferromagnetic chain}, \\
\href{http://dx.doi.org/10.1016/0003-4916(61)90115-4}{Ann. Physics {\bf 16} (1961) 407}.

\bibitem{LM62}
E. H. Lieb and D. Mattis,
\emph{Ordering energy levels of interacting spin systems}, \\
\href{http://dx.doi.org/10.1063/1.1724276}{J. Math. Phys. {\bf 3} (1962), 749}.

\bibitem{Lieb89}
E. H.~Lieb,
\emph{Two theorems on the Hubbard model}, \\
\href{http://dx.doi.org/10.1103/PhysRevLett.62.1201}{Phys. Rev. Lett. {\bf 62} (1989) 1201}.

 \bibitem{Ueda92}
 K. Ueda, H. Tsunetsugu, and M. Sigrist,
 \emph{Singlet ground state of the periodic Anderson model at half filling: A rigorous result}, \\
\href{https://doi.org/10.1103/PhysRevLett.68.1030}{Phys. Rev. Lett. {\bf 68} (1992) 1030}.

\bibitem{Amb92}
T. Xiang and N. d'Ambrumenil,
\emph{Energy-level ordering in the one-dimensional t--J model: A rigorous result}, \\
\href{http://dx.doi.org/10.1103/PhysRevB.46.599}{Phys. Rev. B {\bf 46} (1992) 599}; 
\\[2.5pt]
\emph{Theorem on the one-dimensional interacting-electron system on a lattice}, \\
\href{http://dx.doi.org/10.1103/PhysRevB.46.11179}{Phys. Rev. B {\bf 46}  (1992) 11179}.

\bibitem{AL86}
I. Affleck and E. H. Lieb,
\emph{A proof of part of Haldane's conjecture on spin chains}, \\
\href{http://dx.doi.org/10.1007/BF00400304}{Lett. Math. Phys. {\bf 12} (1986) 57}.

\bibitem{sorella96}
A. Angelucci and S. Sorella,
\emph{Some exact results for the multicomponent t--J model}, \\
\href{http://dx.doi.org/10.1103/PhysRevB.54.R12657}%
{Phys. Rev. B {\bf 54}  (1996) R12657(R)},
\aref{http://arxiv.org/abs/cond-mat/9609107}{cond-mat/9609107}.

\bibitem{Li01}
Y.-Q. Li,
\emph{Rigorous results for a hierarchy of generalized Heisenberg models}, \\
\href{https://doi.org/10.1103/PhysRevLett.87.127208}%
{Phys. Rev. Lett. {\bf 87} (2001) 127208},
\aref{http://arxiv.org/abs/cond-mat/0201060}{cond-mat/0201060}.

\bibitem{H04}
T. Hakobyan,
\emph{The ordering of energy levels for $SU(n)$ symmetric antiferromagnetic chains}, \\
\href{http://dx.doi.org/10.1016/j.nuclphysb.2004.07.032}%
{Nucl. Phys. B {\bf 699} (2004) 575},
\aref{http://arxiv.org/abs/cond-mat/0403587}{cond-mat/0403587}.

\bibitem{Li04}
Y.-Q. Li,   G.-S. Tian, M. Ma, and H.-Q. Lin,
\emph{Ground state and excitations of a four-component fermion model}, \\
\href{http://dx.doi.org/10.1103/PhysRevB.70.233105}%
{Phys. Rev. B {\bf 70} (2004) 233105},
\aref{http://arxiv.org/abs/cond-mat/0407601}{cond-mat/0407601}.

\bibitem{H10}
T. Hakobyan,
\emph{Ordering of energy levels for extended SU(N) Hubbard chain}, \\
\href{http://dx.doi.org/10.3842/SIGMA.2010.024}
 {SIGMA {\bf 6} (2010) 024},
\aref{http://arxiv.org/abs/1003.2147}{arXiv:1003.2147}.

\bibitem{Li98}
Y. Q. Li, M. Ma, D. N. Shi, and  F. C. Zhang,
\emph{$SU(4)$ theory for spin systems with orbital degeneracy}, \\
\href{https://doi.org/10.1103/PhysRevLett.80.3527}%
{Phys. Rev. Lett. {\bf 80} (1998) 3527},
\aref{hhttps://arxiv.org/abs/cond-mat/9804157}{cond-mat/9804157}.


\bibitem{Zhang}
S.-C. Zhang,
\emph{A unified theory based on SO(5) symmetry of superconductivity and antiferromagnetism}, \\
\href{http://dx.doi.org/10.1126/science.275.5303.1089}{Science  {\bf  275}  (1997) 1089}.

\bibitem{so5}
E. Demler, W. Hanke, and S.-C. Zhang,
\emph{SO(5) theory of antiferromagnetism and superconductivity}, \\
\href{https://doi.org/10.1103/RevModPhys.76.909}%
{Rev. Mod. Phys. {\bf 76} (2004) 909},
\aref{https://arxiv.org/abs/cond-mat/0405038}{cond-mat/0405038}.


\bibitem{Honercamp04}
C. Honerkamp and W. Hofstetter,
\emph{Ultracold fermions and the $SU(N)$ Hubbard model}, \\
\href{http://dx.doi.org/10.1103/PhysRevLett.92.170403}{Phys. Rev. Lett. {\bf 92} (2004) 170403},
\aref{http://arxiv.org/abs/cond-mat/0309374}{cond-mat/0309374}; 
\\[2.5pt]
A. V. Gorshkov, M. Hermele, V. Gurarie, C. Xu, P.S.~Julienne, J.~Ye, P.~Zoller, E.~Demler, M. D.~Lukin, and A. M.~Rey,
\emph{Two-orbital SU(N) magnetism with ultracold alkaline-earth atoms}, \\
\href{https://doi.org/10.1038/nphys1535}%
{Nature Phys.  {\bf 6} (2010) 289},
\aref{http://arxiv.org/abs/0905.2610}{arXiv:0905.2610}.
%

\bibitem{rey14}
 M. A. Cazalilla and  A. M. Rey,
\emph{Ultracold Fermi gases with emergent SU(N) symmetry},\\
\href{https://doi.org/10.1088/0034-4885/77/12/124401}%
{Rep. Prog. Phys. \textbf{77} (2014) 124401},
\aref{http://arxiv.org/abs/1403.2792}{arXiv:1403.2792}.



\bibitem{inter-maj}
A. Rahmani and M. Franz,
\emph{Interacting Majorana fermions},\\
\href{https://doi.org/10.1088/1361-6633/ab28ef}%
{Rep. Prog. Phys. {\bf 82} (2019) 084501},
\aref{https://arxiv.org/abs/1811.02593}{arXiv:1811.02593};
\\[2.5pt]
Ching-Kai Chiu, D. I. Pikulin, and M. Franz,
\emph{Strongly interacting Majorana fermions},\\
\href{https://doi.org/10.1103/PhysRevB.91.165402}%
{Phys. Rev. B {\bf 91}, 165402 (2015)},
\aref{https://arxiv.org/abs/1411.5802}{arXiv:1411.5802}.

\bibitem{kitaev01}
A. Kitaev,
\emph{Unpaired Majorana fermions in quantum wires},\\
\href{https://doi.org/10.1070/1063-7869/44/10S/S29}%
{Phys. Usp.  {\bf 44} (2001) 131},
\aref{https://arxiv.org/abs/cond-mat/0010440}{cond-mat/0010440}.



 \bibitem{H15}
 T. Hakobyan,
 \emph{Lowest-energy states in parity-trans\-formation eigenspaces of SO(N) spin chain},\\
\href{https://doi.org/10.1016/j.nuclphysb.2015.07.003}%
{Nucl. Phys. B {\bf 898} (2015) 248},
 \aref{https://arxiv.org/abs/1412.8177}{arXiv:1412.8177}.


\bibitem{Hamermesh}
M. Hamermesh,
\emph{Group Theory and its Application to Physical Problems},
(Dover, New York, 1989).

\bibitem{wil09}
F. Wilczek,
\emph{Majorana returns},\\
\href{http://dx.doi.org/10.1038/nphys1380}{Nature Physics {\bf 5} (2009) 614}.

\bibitem{kitaev11}
L. Fidkowski and A. Kitaev,
\emph{Topological phases of fermions in one dimension},\\
\href{https://doi.org/10.1103/PhysRevB.83.075103}%
{Phys. Rev. B {\bf 83} (2011) 075103},
\aref{https://arxiv.org/abs/1008.4138}{arXiv:1008.4138}.

 \bibitem{MC}
 Z.-X. Li, Y.-F. Jiang, and H. Yao,
\emph{Solving fermion sign problem in quantum Monte Carlo by Majorana representation},\\
\href{https://doi.org/10.1103/PhysRevB.91.241117}%
{Phys. Rev. B {\bf 91} (2015) 241117},
 \aref{https://arxiv.org/abs/1408.2269}{arXiv:1408.2269};
\\[2.5pt]
  L. Wang, Y.-H. Liu, M. Iazzi, M. Troyer, and  G. Harcos,
\emph{Split orthogonal group: A guiding principle for sign-problem-free fermionic simulations},\\
\href{https://doi.org/10.1103/PhysRevLett.115.250601}%
{Phys. Rev. Lett. 115, 250601 (2015)},
\aref{https://arxiv.org/abs/1506.05349}{arXiv:1506.05349};
\\[2.5pt]
  Z. C.~Wei, C. Wu, Y. Li, S. Zhang, and T. Xiang,
 \emph{Majorana positivity and the fermion sign problem of quantum Monte Carlo simulations},\\
\href{https://doi.org/10.1103/PhysRevLett.116.250601}%
{\emph{Phys. Rev. Lett.} \textbf{116} (2016) 250601},
 \aref{https://arxiv.org/abs/1601.01994}{arXiv:1601.01994};
\\[2.5pt]
  \emph{Majorana-time-reversal symmetries: a fundamental principle for sign-problem-free
 quantum Monte Carlo simulations},\\
 \href{https://doi.org/10.1103/PhysRevLett.117.267002}%
 {\emph{Phys. Rev. Lett.} {\bf 117} (2016) 267002},
 \aref{https://arxiv.org/abs/1601.05780}{arXiv:1601.05780}.


\bibitem{Wei15}
Z.-C. Wei, X.-J. Han, Z.-Yu. Xie, and T. Xiang,
\emph{Ground state degeneracy of interacting spinless fermions},\\
\href{https://doi.org/10.1103/PhysRevB.92.161105}%
{Phys. Rev. B \textbf{92} (2015) 161105},
\aref{https://arxiv.org/abs/1412.1578}{arXiv:1412.1578}.


\bibitem{fnote}
Anywhere in  Sec.~\ref{sub:L=2} and  the Appendix, we omit the superscript $L=2$
 over the  subspaces,
 so ${\cal V}_m={\cal V}^2_m$, $V=V^2$, etc.

\bibitem{harada14}
K. Okunishi and K. Harada,
\emph{Symmetry-protected topological order and negative-sign problem for SO(N)
bilinear-biquadratic chains},\\
\href{https://doi.org/10.1103/PhysRevB.89.134422}%
{Phys. Rev. B {\bf 89}  (2014) 134422},
\aref{http://arxiv.org/abs/1312.2643v2}{arXiv:1312.2643}.


\bibitem{super}
T. H. Hsieh, G. B. Hal\'asz, and T. Grover,
\emph{All Majorana models with translation symmetry are supersymmetric},\\
\href{https://doi.org/10.1103/PhysRevLett.117.166802}%
{Phys. Rev. Lett. {\bf 117} (2016) 166802},
\aref{https://arxiv.org/abs/1604.08591}{arXiv:1604.08591}.

 \bibitem{resh}
N. Y. Reshetikhin,
\emph{A method of functional equations in the theory of exactly solvable quantum systems},\\
 \href{http://dx.doi.org/10.1007/BF00400435}{Lett. Math. Phys. {\bf 7} (1983) 205};
\\[2.5pt]
\emph{Integrable models of quantum one-dimensional magnets with $O(n)$ and $Sp(2k)$ symmetry},\\
\href{http://dx.doi.org/10.1007/BF01017501}{Theor. Math. Phys. {\bf 63} (1985) 555}.


\end{thebibliography}
\end{document}